\documentclass{article}
\usepackage{amssymb,amsmath}
\usepackage{graphics}
\usepackage[driver]{graphicx}

\if@twoside  m
\else
\fi \topmargin 5mm \headheight 0mm \headsep 0mm \textheight
225truemm \textwidth 150truemm
\renewcommand{\theequation}{\arabic{section}.\arabic{equation}}
\renewcommand{\thesection}{\Roman{section}}

\newcommand{\D}{{\rm\ d}}                                             
\newtheorem{thm}{Theorem}                          
\newtheorem{lemma}{Lemma}                          
\newtheorem{rem}{Remark}                           

\newcommand{\qed}
           {\mbox{\quad\rule[-1.5pt]{.4em}{1.5ex}}}       
\newcommand{\ie}{{\em i.e.}}
\newcommand{\eg}{{\em e.g.}}

\newcommand{\eps}{\varepsilon}

\begin{document}

\title{Bound states in straight quantum waveguides with combined boundary conditions}
\author{J.~Dittrich$^{a,b}$ and J.~K\v{r}\'\i\v{z}$^{a,c}$}
\date{}
\maketitle
\begin{quote}
{\small \em a) Nuclear Physics Institute, \\ \phantom{e)x}Academy
of Sciences of the Czech Republic, 250 68 \v Re\v z, Czech
Republic (mail address)\\
 b) Doppler Institute of Mathematical Physics,\\
\phantom{e)x}Faculty of Nuclear Sciences and Physical Engineering, Czech
 Technical University,\\
\phantom{e)x}B\v{r}ehov{\'a} 7, 115 19 Prague 1, Czech Republic \\
 c) Faculty of Mathematics and Physics, Charles University,\\
\phantom{e)x}V~Hole\v{s}ovi\v{c}k\'ach~2, 180 00 Prague 8, Czech
Republic
\\
 \rm \phantom{e)x}dittrich@ujf.cas.cz, kriz@ujf.cas.cz}
\vspace{8mm}

\noindent {\small We investigate the discrete spectrum of the
Hamiltonian describing a quantum particle
living in the two-dimensional straight strip.
We impose the combined Dirichlet and Neumann boundary conditions
on different parts of the boundary. 
Several statements on the existence or the absence of the discrete
spectrum are proven for two models with combined boundary
conditions.
Examples of eigenfunctions and eigenvalues are computed
numerically.}
\end{quote}


\section{Introduction}

Quantum waveguides with Dirichlet boundary conditions were
extensively studied (\eg\ \cite{ES}, \cite{GJ}, \cite{DE}, \cite{ESTV},
\cite{Popov}, \cite{Hurt} and references therein).
Their spectral properties essentially depends on the geometry of
the waveguide, especially the existence of bound states induced by
curvature \cite{ES}, \cite{GJ}, \cite{DE} or by coupling of straight waveguides
through windows \cite {ESTV},\cite{Popov} were shown.
The waveguides with Neumann boundary condition were also
investigated in several papers (\eg\ \cite{NSN}, \cite{DP}).
The possible next generalization are waveguides with combined
Dirichlet and Neumann boundary conditions on different parts of 
the boundary. Some very simple combinations of these conditions
appear due to the symmetry of special configurations in systems
studied \eg\ in \cite{ESTV}, \cite{Popov} and \cite{DP}. Such
``combined" systems might be also of interest directly in
nanoscopic physics if interphases modelled by different conditions
could be realized.
The presence of different boundary
conditions also gives rise to nontrivial spectral properties like
existence of bound states.

In the present paper, we consider two simple cases of straight
planar waveguide of constant width with combined boundary conditions.
We show the examples with and without the presence of bound
states.
The systems we are going to study are sketched on Fig.~1.
We consider a Schr\"odinger particle whose motion is confined to a planar
strip of width $d$.
For definiteness we assume that it is placed to the upper side
of the $x-$axis.
On the part of the boundary the Neumann condition is imposed (thin lines in the
picture),
while on the other part the Dirichlet one holds (thick lines). 
The length of the overlay of Neumann boundaries is $2\delta$ and it is
placed to both sides of $y-$axis in both cases.
We shall denote this configuration space by $\Omega=\mathbb{R} \times
(0,d)$ and its particular parts by $\Omega_I=(-\infty,-\delta)\times
(0,d)$, $\Omega_{II}=(-\delta,\delta)\times (0,d)$ and $\Omega_{III}=
(\delta,\infty)\times (0,d)$.
As we are going to prove several statements that are valid
for more general combination of boundary conditions, let us define
several objects.
Let there is a finite number of points on the boundary $\partial \Omega$,
where boundary condition is changing, which we denote $P_k=\langle
x_k,y_k\rangle, k=1,\ldots,M$.
We can choose the numbering so as $y_k=d$ for $k=1,\ldots,M'$ and
$x_1<x_2<\ldots<x_{M'}$ and $y_k=0$ for $k=M'+1,\ldots,M$ and
$x_{M'+1}<x_{M'+2}<\ldots<x_M$.
Let $\partial \Omega=\mathcal{D}\cup\mathcal{N}\cup
\bigcup_{k=1}^M\{P_k\}$, where $\mathcal{D}$ is a union of finite
number of intervals in $\partial \Omega$, where Dirichlet condition
is imposed and $\mathcal{N}=\bigl(\partial
\Omega \setminus \mathcal{D}\bigr)^0$ is similar for Neumann
condition.
For our examples we have
\begin{itemize}
\item[\textrm{A)}]
\begin{eqnarray}
\nonumber
\mathcal{D}&= & \Bigl{\{}\langle x,0\rangle | x<-\delta
\Bigr{\}} \cup \Bigl{\{}\langle x,d\rangle | x>\delta
\Bigr{\}}\\
\nonumber
\mathcal{N}&= & \Bigl{\{}\langle x,0\rangle | x>-\delta
\Bigr{\}} \cup \Bigl{\{}\langle x,d\rangle | x<\delta
\Bigr{\}}\\
\nonumber
P_1&= &\langle \delta,d\rangle\ ;\ P_2=\langle -\delta,0\rangle
\end{eqnarray}
\item[\textrm{B)}]
\begin{eqnarray}
\nonumber
\mathcal{D}&= & \Bigl{\{}\langle x,d\rangle | (x<-\delta) \vee
 (x>\delta)\Bigr{\}}\\
\nonumber
\mathcal{N}&= & \Bigl{\{}\langle x,0\rangle | x \in \mathbb{R}
\Bigr{\}} \cup \Bigl{\{}\langle x,d\rangle | -\delta<x<\delta
\Bigr{\}}\\
\nonumber
P_1&= &\langle -\delta,d\rangle\ ;\ P_2=\langle \delta,d\rangle
\end{eqnarray}
\end{itemize}

In the next section we define the Hamiltonian as Laplace operator
with chosen boundary conditions with the help of a quadratic form.
We also explicitly give the operator domain which is larger than
the Sobolev space $H^2(\Omega)$.
Due to this fact the proof of its form is a little complicated.
In the section III we study the question of bound state existence
below the treshold of essential spectrum.
The proved results are illustrated in Section IV by numerical
calculations.
Some technical points are left to Appendices.

\begin{figure}
   \begin{picture}(120,160)
      \linethickness{1pt}
      \put(30,180){\line(1,0){215}}
      \put(165,140){\line(1,0){215}}
      \linethickness{3pt}
      \put(30,140){\line(1,0){135}}
      \put(245,180){\line(1,0){135}}
      \thinlines
      \put(205,121){\vector(0,1){70}}
      \put(165,125){\vector(1,0){80}}
      \put(245,125){\vector(-1,0){80}}
      \put(380,140){\vector(1,0){20}}
      \put(10,140){\line(1,0){20}}
      \put(35,140){\vector(0,1){40}}
      \put(165,140){\line(0,-1){17}}
      \put(245,140){\line(0,-1){17}}
      \put(35,155){\vector(0,-1){15}}
      \put(20,155){$d$}
      \put(210,191){y}
      \put(395,130){x}
      \multiput(165,140)(0,9){5}{\line(0,1){4}}
      \multiput(245,140)(0,9){5}{\line(0,1){4}}
      \put(215,127){$2\delta$}
      \put(5,200){\bf{A)}}
      \put(100,155){I}
      \put(210,155){II}
      \put(320,155){III}
      \linethickness{1pt}
      \put(165,60){\line(1,0){80}}
      \put(30,20){\line(1,0){350}}
      \linethickness{3pt}
      \put(30,60){\line(1,0){135}}
      \put(245,60){\line(1,0){135}}
      \thinlines
      \put(205,1){\vector(0,1){70}}
      \put(165,5){\vector(1,0){80}}
      \put(245,5){\vector(-1,0){80}}
      \put(380,20){\vector(1,0){20}}
      \put(10,20){\line(1,0){20}}
      \put(35,20){\vector(0,1){40}}
      \put(165,20){\line(0,-1){17}}
      \put(245,20){\line(0,-1){17}}
      \put(35,35){\vector(0,-1){15}}
      \put(20,35){$d$}
      \put(210,71){y}
      \put(395,10){x}
      \multiput(165,20)(0,9){5}{\line(0,1){4}}
      \multiput(245,20)(0,9){5}{\line(0,1){4}}
      \put(215,7){$2\delta$}
      \put(5,90){\bf{B)}}
      \put(100,35){I}
      \put(210,35){II}
      \put(320,35){III}
   \end{picture}

\vspace{5mm}

\caption{Straight quantum waveguides with combined boundary conditions. The
thin
 lines denote the Neumann boundary condition, the thick lines the Dirichlet one. }
   \end{figure}
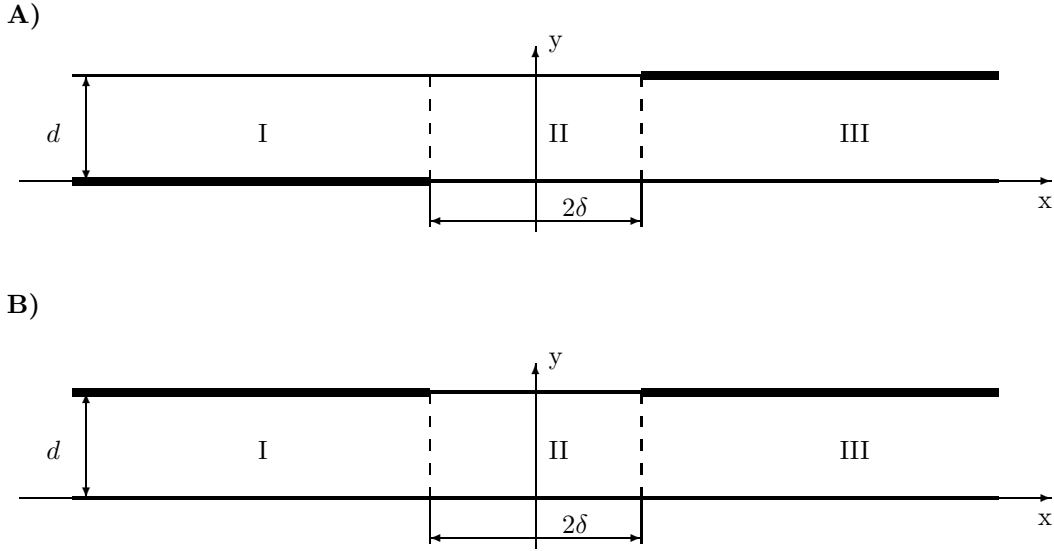

\section{The Hamiltonian}

Putting $\hbar^2 /2m=1$, we may identify the particle Hamiltonian
with the self-adjoint operator on the Hilbert space $L^2(\Omega)$,
defined in the following way. Let us define a quadratic form
\begin{equation}
q_0(f,g)=\int_{\Omega} \overline{\nabla f}\cdot \nabla g \D^2 x
\rm {\ \ with\ domain\ } Q(q_0)=\bigl {\{} f\in H^1(\Omega) | f\upharpoonright
\mathcal{D}=0 \bigr{\}},
\end{equation}
where $H^1=\bigl {\{} f \in L^2(\Omega)| \nabla f \in L^2(\Omega) \bigr {\}}$
is the standard Sobolev space and we denote as $f\upharpoonright
\mathcal{D}$ the trace of function $f$ on $\mathcal{D}$.
Now $q_0$ is obviously densely defined, symmetric and below bounded quadratic form.
The form $q_0$ is also closed as a direct consequence of Theorem 7.53 in
\cite{Adams}.

There is the unique self-adjoint operator associated with this
form (see \eg\ \cite{BEH}, Theorem 4.6.8).
We denote this operator $-\triangle^{\Omega}_{DN}$ and its domain
$D(\Omega)$.
It is our Hamiltonian.
We will show, that this operator acts as the usual Laplace
operator with the Dirichlet condition on $\mathcal{D}$ and Neumann
condition on $\mathcal{N}$. 
\begin{thm}
\label{Operator Domain}
The domain of the operator $-\triangle^{\Omega}_{DN}$
is
\begin{eqnarray}
\label{Operator Domain 2}
D(\Omega)&=&\bigl {\{}f \in H^1(\Omega) | -\triangle f \in
L^2(\Omega), f\upharpoonright \mathcal{D}=0, {\partial f \over \partial y}\upharpoonright
\mathcal{N}=0\bigr {\}},\\
\nonumber
-\triangle^{\Omega}_{DN}f&=&-\triangle f {\rm \ \ for \ every  \ \ } f \in
D(\Omega).
\end{eqnarray}
\end{thm}
\begin{PF}
First, we know that $D(\Omega) \subset Q(q_0)$. Moreover $f\in
D(\Omega)$ if and only if there exists a function $h \in
L^2(\Omega)$ such that for all $g \in Q(q_0)$ the equality
$q_0(g,f)=\bigl (g,h \bigr )_{L^2(\Omega)}$ holds.
Then $h=-\triangle^{\Omega}_{DN}f$ (see \cite{BEH}, Theorem 4.6.8).
Let $g$ be any function from $C_0^{\infty}(\Omega)$.
Then $g \in Q(q_0)$ and $\bigl (\nabla g,\nabla f \bigr )_{L^2(\Omega)}
=\bigl (g,-\triangle f \bigr )_{L^2(\Omega)}$ using only
definition of the distributional derivatives.
So $-\triangle^{\Omega}_{DN}f=-\triangle f$ for all $f \in
D(\Omega)$.
We know now
\begin{equation}
\label{Operator Domain 1}
D(\Omega)=\Bigl {\{} f \in Q(q_0) | -\triangle f \in L^2(\Omega),
\Bigl(\forall g \in Q(q_0)\Bigr )\,\Bigl ( \bigl(g,-\triangle f \bigr )_{L^2(\Omega)}
=\bigl (\nabla g,\nabla f\bigr )_{L^2(\Omega)}\Bigr ) \Bigr{\}}
\end{equation}

Now we prove the implication $f \in D(\Omega) \Rightarrow {\partial f \over
\partial y}\upharpoonright \mathcal{N} = 0$.
Let $\Omega_0$ be an open subset of $\Omega$, such that $P_k
\notin \overline{\Omega}_0, k=1,\ldots,M$.
We will show that every $f \in D(\Omega)$ belongs to
$H^2(\Omega_0)$ for such subdomains.
There exist real, positive numbers $\eta_k$, $k=1,\ldots,M$ such that the open balls
$B(P_k,2\eta_k)$ have empty intersections with $\Omega_0$. We can
choose $\eta_k$ so small, that even $\overline{B(P_k,2\eta_k)}\cap
\overline {B(P_{k'},2\eta_{k'})} = \emptyset$ for $k \neq k'$.
Let we denote
\begin{eqnarray}
\nonumber
\omega_0 &=& \bigl (-\infty,x_1-\eta_1\bigr ) \times
(d,2d)\\
\nonumber
\vdots\\
\nonumber
\omega_{M'-1} &=& \bigl (x_{M'-1}+\eta_{M'-1} ,x_{M'}-\eta_{M'} \bigr ) \times
(d,2d)\\
\nonumber
\omega_{M'} &=& \bigl (x_{M'}+\eta_{M'} ,\infty\bigr ) \times
(d,2d)\\
\nonumber
\omega_{M'+1} &=& \bigl (-\infty,x_{M'+1}-\eta_{M'+1} \bigr ) \times
(-d,0)\\
\nonumber
\vdots\\
\nonumber
\omega_{M+1} &=& \bigl (x_M+\eta_M ,\infty\bigr ) \times
(-d,0)
\end{eqnarray}
and let $\tilde\omega_k$ be the reflection of $\omega_k$ to the
domain $\Omega$ for $k=0,\ldots, M+1$, \ie\ $\tilde
\omega_k=\mathcal{R}_1(\omega_k)$ for $k=0,\ldots, M'$ and $\tilde
\omega_k=\mathcal{R}_2(\omega_k)$ for $k=M'+1,\ldots, M+1$, where the
bijections $\mathcal{R}_i:\mathbb{R}^2\rightarrow\mathbb{R}^2$, $i=1,2$ are defined
as follows: $\mathcal{R}_1(\langle x,y\rangle)=\langle
x,2d-y\rangle$ and $\mathcal{R}_2(\langle x,y\rangle)=\langle
x,-y\rangle$.
Let $\tilde \Omega=\Bigl(\overline{\bigcup_{k=0}^{M+1}\omega_k \cup
\Omega}\Bigr)^0$.
In fact, we can say that $\tilde{\Omega}$ is the original domain
$\Omega$ with its copy on each side of its boundary, from which we
cut the columns $[x_k-\eta_k,x_k+\eta_k]\times [d,2d)$, resp. $[x_k-\eta_k,x_k+\eta_k]\times
(-d,0]$, above, resp. below, the point $P_k$ depending on which part of
$\partial \Omega$ the point $P_k$ lies.
We construct a function $\tilde f \in L^2(\tilde {\Omega})$
as follows.
For every point $\langle x,y \rangle \in \tilde {\Omega} \setminus
\partial \Omega$ we define (so we see that the function will be
defined almost everywhere in $\tilde \Omega$):
$$
\tilde f(x,y) = \left \{
\begin{array}{lcll}
f(x,y) & {\rm for} & \langle x,y\rangle \in \Omega\\
\\
f(x,2d-y) & {\rm for} & \langle x,2d-y\rangle \in \Omega, & \langle x,d\rangle \in \mathcal{N}\\
\\
-f(x,2d-y) & {\rm for} & \langle x,2d-y\rangle \in \Omega, & \langle x,d\rangle \in \mathcal{D}\\
\\
f(x,-y) & {\rm for} & \langle x,-y\rangle \in \Omega, & \langle x,0\rangle \in \mathcal{N}\\
\\
-f(x,-y) & {\rm for} & \langle x,-y\rangle \in \Omega, & \langle x,0\rangle \in \mathcal{D}.\\
\end{array} \right .
$$
Now for any $\varphi \in C_0^\infty(\tilde \Omega)$ we can write
\begin{eqnarray}
\nonumber
&{\rm \ } & \bigl (-\triangle \tilde f,\varphi \bigr )_{L^2(\tilde
\Omega)} = (\tilde f,-\triangle\varphi \bigr )_{L^2(\tilde
\Omega)} = (f,-\triangle\varphi \bigr )_{L^2(\Omega)} +
\sum_{k=0}^{M+1}(\tilde f,-\triangle\varphi \bigr
)_{L^2(\omega_k)}=\\
\nonumber
&=&(f,-\triangle\varphi \bigr )_{L^2(\Omega)} \mp
\sum_{k=0}^{M'}(-1)^k\int_{\tilde\omega_k}
\overline {f(x,y)}\triangle\varphi(x,2d-y) \D^2x\\
\label{pokracovani}
&+&(-1)^s \sum_{k=M'+1}^{M+1}(-1)^k\int_{\tilde \omega_k}
\overline{f(x,y)}\triangle\varphi(x,-y) \D^2x
=(f,-\triangle\tilde\varphi \bigr )_{L^2(\Omega)},
\end{eqnarray}
where we used the definition of the distributional derivatives and
the substitution $y=2d-y$, resp. $y=-y$.
The sign $-$ in $\mp$ is valid for these systems, where Neumann
condition is imposed on $\bigl \{ \langle x,d\rangle |x 
\in(-\infty,x_1)\bigr\}$, the sign $+$ for others.
The number $s$ equals $0$ or $1$, so as $s+M'+1$ is odd for systems,
where Neumann condition is imposed on  $\bigl \{ \langle x,0\rangle
|x 
\in(-\infty,x_{M'+1})\bigr \}$ and even for others.
Finally, new function is defined on the domain $\Omega$ as
$$
\tilde
\varphi(x,y)=\varphi(x,y)\pm\sum_{k=0}^{M'}(-1)^k
\chi_{\tilde\omega_k}(x,y)\varphi(x,2d-y)-(-1)^s
\sum_{k=M'+1}^{M+1}(-1)^k \chi_{\tilde\omega_k}(x,y)\varphi(x,-y),
$$
where $\chi_\omega$ is the standard characteristic function of the set
$\omega$.
Taking into consideration the construction of the domain $\tilde \Omega$ and
that ${\rm supp\ \ }\varphi \subset \tilde \Omega$, we conclude that
$\tilde\varphi \in C^\infty(\overline\Omega)$ and it has a bounded support.
Further, we know that the trace $\tilde\varphi\upharpoonright \partial
\Omega$ equals
$\lim_{y\rightarrow d^-}\tilde\varphi(x,y)$ in the point $\langle x,d\rangle$ and similarly
$\lim_{y\rightarrow 0^+}\tilde\varphi(x,y)$ in the point $\langle x,0\rangle$ for smooth functions (see the
definition of traces, \eg\ in \cite{Adams}).
So the traces on $\mathcal{D}\cap \tilde\Omega$ are
\begin{eqnarray}
\nonumber
(\tilde\varphi \upharpoonright \partial
\Omega)(x,d)&= &\lim_{y\rightarrow d^-} \tilde \varphi (x,y) = \lim_{y\rightarrow
d^-}\bigl ( \varphi(x,y)-\varphi(x,2d-y) \bigr)=0\\
\nonumber
(\tilde\varphi \upharpoonright \partial
\Omega)(x,0)&= &\lim_{y\rightarrow 0^+} \tilde \varphi (x,y) = \lim_{y\rightarrow
0^+}\bigl ( \varphi(x,y)-\varphi(x,-y) \bigr)=0.
\end{eqnarray}
In a similar way for parts of boundary $\partial\Omega$ with Neumann
condition inside $\tilde\Omega$
\begin{eqnarray}
\nonumber
\bigl ({\partial\tilde\varphi \over \partial y}\upharpoonright \partial
\Omega\bigr )(x,d)&= &\lim_{y\rightarrow d^-} {\partial\tilde \varphi \over \partial y}
 (x,y) = \lim_{y\rightarrow
d^-}{\rm\ }{\partial \over \partial y}\bigl ( \varphi(x,y)+\varphi(x,2d-y) \bigr)=0\\
\nonumber
\bigl ({\partial\tilde\varphi \over \partial y}\upharpoonright \partial
\Omega\bigr )(x,0)&= &\lim_{y\rightarrow 0^+} {\partial\tilde \varphi \over \partial y}
 (x,y) = \lim_{y\rightarrow
0^+}{\rm\ }{\partial \over \partial y}\bigl ( \varphi(x,y)+\varphi(x,-y)
\bigr)=0.
\end{eqnarray}
On the rest of the boundary $\partial \Omega$, \ie\ $\partial
\Omega \setminus \tilde\Omega$ both Dirichlet and Neumann
conditions are satisfied, what can be seen from the definition of
$\tilde\varphi$.
So it is clear that $\tilde \varphi \in H^2(\Omega)$ and it
satisfies right boundary conditions.
It is easy to check, that all such functions belong to
$D(\Omega)$, using the Gauss Theorem.
Because both functions $f$ and $\tilde \varphi$ are in
$D(\Omega)$, which is a subset of $Q(q_0)$, we can continue the
calculation from (\ref{pokracovani}).
\begin{eqnarray}
\nonumber
&{\rm\ }&  \bigl (f,-\triangle\tilde\varphi \bigr )_{L^2(
\Omega)}= \bigl (\nabla f,\nabla\tilde\varphi \bigr )_{L^2(
\Omega)}= \bigl (-\triangle f,\tilde \varphi \bigr )_{L^2(
\Omega)}= \bigl (-\triangle f,\varphi \bigr )_{L^2(
\Omega)}\mp\\
\nonumber
&\mp& \sum_{k=0}^{M'}(-1)^k\int_{\omega_k}
\overline {\triangle f(x,2d-y)}\varphi(x,y) \D^2x
+(-1)^s \sum_{k=M'+1}^{M+1}(-1)^k\int_{\omega_k}
\overline{\triangle f(x,-y)}\varphi(x,y) \D^2x=\\
\nonumber
&=& \bigl (F,\varphi \bigr )_{L^2(\tilde
\Omega)},
\end{eqnarray}
where we used the ``reflection" substitution again and $F$ is a
function defined by the last formula.
Here $F \in L^2(\tilde \Omega)$, because it is the sum of the
finite number of $L^2$-functions.
As we choose the function $\varphi$ arbitrarily, we see
that $-\triangle \tilde f=F \in L^2(\tilde \Omega)$.
Let $\psi \in C^\infty( \mathbb{R}^2)$, it is bounded together
with its first and second derivatives and let supp\ $\psi \subset
\tilde\Omega$.
Then $\psi\tilde f \in L^2( \mathbb{R}^2)$.
Using Leibnitz rule and several times a lemma from the section IX.6
in \cite{RS}, we conclude that even $-\triangle(\psi\tilde f)
\in L^2( \mathbb{R}^2)$
(Leibnitz rule itself does not give the result unless we know
$\nabla\tilde f\in L^2(\tilde \Omega)$ ).
We now use this lemma once more and we get the result, that
$\psi\tilde f\in H^2(\mathbb{R}^2)$.
We can choose a function $\psi$ so as $\psi\upharpoonright
\Omega_0=1$.
It is possible, because $\Omega_0 \subset\subset \tilde \Omega$
and our regions have a simple form at $x\rightarrow\pm\infty$.
Let $r_1<r_2<r_3$ be the real positive numbers such that $[x_k-2\eta_k,x_k+2\eta_k]\in(-r_1,r_1)$ for every
$k=1,\ldots,M$ and denote
 $R_2=(-r_2,r_2)\times (0,d)$, $R_3=(-r_3,r_3) \times
 (-d,2d)$.
Then using \cite{DSch}, Lemma XIV.2.1 we find a function
$\psi_1\in C_0^\infty(\tilde\Omega\cap R_3)$, such that $0\leq\psi_1(x)\leq
1$ for all $x \in \tilde\Omega\cap R_3$ and $\psi_1(x)=1$ for
$x\in \overline{\Omega_0\cap R_2}$.
This function has compact support in $\tilde\Omega$, so its
derivatives are bounded.
Let now $\gamma\in C_0^\infty(-d,2d)$, such that $\gamma(y)=1$ for
$y\in(0,d)$ (it can be constructed according to the same lemma as $\psi_1$).
Let $\beta\in C^\infty( \mathbb{R})$ such that
$\beta(x)=1$ for $|x|\geq r_2$, $\beta(x)=0$ on the interval
$[-r_1,r_1]$ (we can again use the same lemma for construction of $1-\beta$).
Then $\psi=\psi_1(1-\beta)+\gamma\beta$ satisfies all desired
properties.
Thus $\tilde f\psi\upharpoonright \Omega_0 =\tilde f
\upharpoonright \Omega_0=f\upharpoonright
\Omega_0$, so $f \in H^2(\Omega_0)$.

Now let us take any interval $(a,b)$, such $x_k \not \in [a,b]$
for $k=1,\ldots,M$.
Let $\xi \in C_0^\infty(a,b)$ be a real function.
Because region $(a,b)\times (0,d)$ satisfies all conditions for
$\Omega_0$, $f \in H^2\bigl( (a,b)\times (0,d) \bigr)$.
Using Leibnitz rule
we can see that $\xi f \in H^2\bigl( (a,b)\times (0,d) \bigr)$.
For any $g \in Q(q_0)$ we have $g\xi \in Q(q_0)$ and
\begin{eqnarray}
\nonumber
\bigl (\nabla g,\nabla (f\xi) \bigr )_{L^2(\Omega)}&=& \bigl
(\nabla g, \xi\nabla f \bigr
)_{L^2(\Omega)} +\Bigl ({\partial g\over \partial x},f {d \xi \over d x} \Bigr
)_{L^2(\Omega)} =\bigl
(\xi\nabla g,\nabla f \bigr
)_{L^2(\Omega)}+\\
\nonumber
&+& \Bigl (g{d \xi\over d x},{\partial f\over \partial x} \Bigr
)_{L^2(\Omega)}
-\Bigl (g{d \xi\over d x},{\partial f\over \partial x} \Bigr
)_{L^2(\Omega)} + \Bigl({\partial g\over\partial x}, f{d \xi \over d x} \Bigr
)_{L^2(\Omega)} = \\
\nonumber
&=&\bigl(g\xi,-\triangle f \bigr
)_{L^2(\Omega)} +\int_\Omega{\partial\bigl(\overline{g(x,y)} f(x,y)\bigr)\over\partial
x}{d \xi (x)\over d x}\D x\D y-\\
\nonumber
&-& 2\Bigl (g{d\xi\over d x},{\partial f\over \partial x} \Bigr
)_{L^2(\Omega)}=
\bigl(g,-\xi\triangle f \bigr
)_{L^2(\Omega)}-2 \Bigl (g,{d \xi\over d x}\cdot{\partial f\over \partial x} \Bigr
)_{L^2(\Omega)}-\\\label{xif}
&-&\Bigl(g,f{d^2\xi\over d x^2}  \Bigr
)_{L^2(\Omega)}=\bigl (g,-\triangle (f\xi)\ \bigr
)_{L^2(\Omega)}.
\end{eqnarray}
Hence $\xi f \in D(\Omega)$.
Using the Gauss Theorem (it can be used for
$H^2$-functions) we get for any $g \in Q(q_0)$
$$
\bigl (g,-\triangle(\xi f)\bigr )_{L^2(\Omega)} =
\bigl (\nabla g,\nabla(\xi f)\bigr )_{L^2(\Omega)}
-\int_a^b\Bigl(\overline{g(x,d)}\xi(x){\partial f \over \partial
y}(x,d)-\overline{g(x,0)}\xi(x){\partial f \over \partial
y}(x,0) \Bigr)\D x
$$
So due to (\ref{xif})
$$
\int_a^b\Bigl(\overline{g(x,d)}\xi(x){\partial f \over \partial
y}(x,d)-\overline{g(x,0)}\xi(x){\partial f \over \partial
y}(x,0) \Bigr)\D x=0
$$
for any considered $a,b$ and any $g \in Q(q_0), \xi \in C_0^\infty(a,b)$.
Now we conclude that
$$
{\partial f \over \partial y}\upharpoonright \mathcal{N}=0 {\rm \ \
a.e.}
$$
This finishes the second part of proof.

It remains to show that if $f$ satisfies all conditions from
(\ref{Operator Domain 2}) then $f\in D(\Omega)$ (in the sense of
definition (\ref{Operator Domain 1})).
Let $\Omega_{0,\eps}=\Omega\setminus\overline{\bigcup_{k=1}^M
B(P_k,\eps)}$, $\Omega_\eps=\bigcup_{k=1}^MB(P_k,\eps)\cap \Omega$.
Because we know from the previous part of the proof that $f \in
H^2(\Omega_{0,\eps})$, we can use the Gauss theorem:
\begin{eqnarray}
\nonumber
&{\rm \ }&\bigl (\nabla g,\nabla f\bigr )_{L^2(\Omega)}+\bigl (g,\triangle f\bigr
)_{L^2(\Omega)} =\\
\nonumber
&=& \bigl (\nabla g,\nabla f\bigr )_{L^2(\Omega_{0,\eps})}+
\bigl (g,\triangle f\bigr )_{L^2(\Omega_{0,\eps})}+\bigl (\nabla g,\nabla f\bigr
)_{L^2(\Omega_\eps)}+\bigl (g,\triangle f\bigr
)_{L^2(\Omega_\eps)}=\\
\label{epsilon rozklad}
&=& -
\sum_{k=1}^M\int_0^\pi{\partial\tilde f_k(\eps,\varphi)
\over \partial r}\overline{\tilde g_k(\eps,\varphi)} \eps \D \varphi +
\bigl (\nabla g,\nabla f\bigr )_{L^2(\Omega_\eps)}+\bigl (g,\triangle f\bigr
)_{L^2(\Omega_\eps)},
\end{eqnarray}
where $\tilde f_k$, $\tilde g_k$ are the transformations of $f$,
$g$ to the polar coordinates in the neighborhood of each $P_k$
in the way, that the region $(0,\eps)\times(0,\pi)\subset\Omega$,
$P_k$ is the origin of polar coordinates and $\tilde f_k$ satisfies
the Dirichlet condition for $\varphi=0$.
We can see that the last two terms in (\ref{epsilon rozklad}) go
to zero as $\eps \rightarrow 0$, because $\nabla f$, $\nabla g$,
$-\triangle f$, $g\in L^2(\Omega)$ and the measure of
$\Omega_\eps$ goes to zero.
So we only have to prove, that
\begin{equation}
\label{povrchovy clen}
\sum_{k=1}^M\int_0^\pi{\partial\tilde f_k(r_n,\varphi)
\over \partial r}\overline{\tilde g_k(r_n,\varphi)} r_n \D \varphi
\rightarrow 0 {\rm \ \ as} {\rm \ \ }n \rightarrow \infty
\end{equation}
for some sequence $\{r_n\}_{n=0}^\infty$, $\lim_{n\rightarrow \infty}
r_n=0$.
We will show that each term in this sum tends to zero.
For simplicity we will not write indices in the following text.
We will decompose $\tilde f$ to the orthonormal transverse basis
which respects our boundary conditions.
\begin{equation}
\label{rozklad f}
\tilde f (r,\varphi) = \sum_{k=0}^\infty \sqrt{2\over \pi} F_k(r)
\sin{2k+1 \over 2}\varphi
\end{equation}
and in the same way
\begin{equation}
\label{rozklad g}
\tilde g (r,\varphi) = \sum_{k=0}^\infty \sqrt{2\over \pi} G_k(r)
\sin{2k+1 \over 2}\varphi.
\end{equation}
Let $R$ be small positive real number, so as $\min_{k\neq k'}
{\rm dist}(P_k,P_{k'})>R$.
It is easy to check the following equivalences:
\begin{eqnarray}
\label{H1 polar}
f,g \in H^1\bigl(\Omega \cap B(P,R)\bigr ) &\Leftrightarrow & \tilde f,
\tilde g, {\partial \tilde f \over\partial r}, {\partial \tilde g \over\partial
r}, {1 \over r}{\partial \tilde f \over\partial \varphi},
{1 \over r}{\partial \tilde g \over\partial \varphi} \in
L^2\bigl((0,R)\times(0,\pi),r \D r \D \varphi\bigr )\\
\nonumber
\\
\label{Laplace polar}
\triangle f \in L^2\bigl(\Omega \cap B(P,R)\bigr ) &\Leftrightarrow
&\Bigl({\partial^2\tilde f \over \partial r^2}+{1 \over r}{\partial \tilde f \over \partial
r}+{1 \over r^2}{\partial^2 \tilde f \over \partial \varphi^2}
\Bigr ) \in L^2\bigl((0,R)\times(0,\pi),r \D r \D \varphi\bigr ).
\end{eqnarray}
Using (\ref{H1 polar}) we can decompose ${1 \over r}{\partial \tilde f \over\partial
\varphi}$ to the orthonormal transverse basis
$$
{1 \over r}{\partial \tilde f \over\partial \varphi}=\sum_{k=0}^\infty \sqrt{2\over \pi} a_k(r)
\cos{2k+1 \over 2}\varphi.
$$
For almost every $r$ we have
$$
a_k(r)=\sqrt{2 \over \pi} \int_0^\pi {1 \over r}{\partial \tilde f \over \partial
\varphi}(r,\varphi) \cos{2k+1 \over 2}\varphi \D\varphi
$$
and
\begin{eqnarray}
\nonumber
F_k(r)&=&\sqrt{2 \over \pi} \int_0^\pi \tilde f(r,\varphi) \sin{2k+1 \over 2}\varphi
\D\varphi=\\
\nonumber
&=&-\sqrt{2 \over \pi} {2 \over 2k+1} \int_0^\pi \tilde f(r,\varphi) {\partial \over
\partial \varphi }\cos{2k+1 \over 2}\varphi \D\varphi=\\
\nonumber
&=&\sqrt{2 \over \pi} {2 \over 2k+1}\int_0^\pi {\partial \tilde f \over \partial
\varphi}(r,\varphi) \cos{2k+1 \over 2}\varphi \D\varphi+\sqrt{2 \over \pi} {2 \over
2k+1} \tilde f(r,0)=\\
\nonumber
&=&{2 \over 2k+1} r a_k(r),
\end{eqnarray}
due to the boundary condition $\tilde f(r,0)=0$.
So
\begin{equation}
\label{ak}
a_k(r)={1\over r}{2k+1 \over 2} F_k(r).
\end{equation}
Now we decompose in the same way ${\partial \tilde f \over
\partial r}$,
$$
{\partial \tilde f \over\partial r}=\sum_{k=0}^\infty \sqrt{2\over \pi} b_k(r)
\sin{2k+1 \over 2}\varphi.
$$
Let $\{\xi_{k,n}\}_{n=1}^\infty$ be a sequence of
$C_0^\infty(0,\pi)$ functions so as
$\lim_{n\rightarrow\infty}\|\xi_{k,n} -
\sqrt{2 \over \pi}\sin{2k+1 \over 2}\varphi\|_{L^2(0,\pi)}=0$ for
$k=1,\ldots,M$ and let $\omega \in C_0^\infty(0,R)$.
Then using twice the definition of the distributional derivatives,
definitions of $F_k$ and $b_k$ and the fact that $\omega\xi_{k,n}\in
C_0^\infty\bigl((0,R)\times(0,\pi)\bigr )$we get
\begin{eqnarray}
\nonumber
&{\rm \ }&\int_0^R{d \over d r}F_k(r)\omega(r) \D r=-\int_0^RF_k(r)\omega'(r) \D
r=\\
\nonumber
&=&-\int_0^R\omega'(r)\int_0^\pi\sqrt{2 \over \pi}\tilde
f(r,\varphi)\sin{2k+1 \over 2}\varphi \D \varphi \D r=\\
\nonumber
&=& -\lim_{n\rightarrow\infty}\int_0^R\int_0^\pi\omega'(r)\xi_{k,n}(\varphi)\tilde f(r,\varphi)\D
\varphi \D r=\lim_{n\rightarrow\infty}\int_0^R\int_0^\pi\omega(r)\xi_{k,n}(\varphi){\partial
\tilde f \over \partial r}(r,\varphi)\D \varphi \D r=\\
\nonumber
&=&\int_0^R\omega(r)\int_0^\pi\sqrt{2 \over \pi}
{\partial \over \partial r}\tilde f(r,\varphi)\sin{2k+1 \over 2}\varphi \D \varphi \D
r=\int_0^R b_k(r)\omega(r) \D r.
\end{eqnarray}
Because $\omega(r)$ was chosen arbitrarily, we conclude that
\begin{equation}
\label{bk}
b_k(r)={dF_k(r) \over d r}=F'_k(r).
\end{equation}
The same procedure we can apply to $\tilde g(r,\varphi)$.
Using(\ref{ak}) and (\ref{bk}) we  know, that the series
(\ref{rozklad f}) and (\ref{rozklad g}) can be differentiated
by terms.
Now we use the similar derivation for $\triangle f$.
Let
$$
\Bigl({\partial^2\tilde f \over \partial r^2}(r,\varphi)+{1 \over r}{\partial \tilde f \over \partial
r}(r,\varphi)+{1 \over r^2}{\partial^2 \tilde f \over \partial
\varphi^2}(r,\varphi)
\Bigr )=\sqrt{2\over\pi}\sum_{k=0}^\infty c_k(r)\sin{2k+1 \over 2}\varphi.
$$
From the first part of the proof we know $\tilde f(r,.)\in
H^2(0,\pi)$ for a.e. $r$ and we can compute
\begin{eqnarray}
\nonumber
&{\rm \ }&\int_0^R\biggl (F''_k(r)+{1\over r}F'_k(r)-{1 \over
r^2}\Bigl ({2k+1 \over 2}\Bigr)^2F_k(r)\biggr)r\omega(r)\D r=
-\int_0^R r F'_k(r)\omega'(r) \D r-\\
\nonumber
&-& \Bigl ({2k+1 \over 2}\Bigr)^2\int_0^R {1\over r} F_k(r)\omega(r) \D r=
\int_0^R\bigl (r\omega'(r)\bigr )'\int_0^\pi \sqrt{2\over\pi}\tilde
f(r,\varphi)\sin{2k+1\over 2}\varphi \D \varphi \D r-\\
\nonumber
&-&\Bigl ({2k+1 \over 2}\Bigr)^2\int_0^R{\omega(r)\over
r}\int_0^\pi\sqrt{2\over\pi}\tilde f(r,\varphi)\sin{2k+1\over 2}\varphi \D
\varphi\D r=\\
\nonumber
&=&\lim_{n\rightarrow\infty}\int_0^R\bigl (r\omega'(r)\bigr
)'\int_0^\pi \tilde f(r,\varphi)\xi_{n,k}(\varphi)\D \varphi \D
r+\\
\nonumber &+&\int_0^R{\omega(r)\over r}\int_0^\pi\sqrt{2\over\pi}\tilde
f(r,\varphi){d^2 \sin{2k+1\over2}\varphi \over d\varphi^2} \D
\varphi \D r=\\
\nonumber
&=&-\lim_{n\rightarrow\infty}\int_0^R\int_0^\pi\omega'(r)
\xi_{n,k}(\varphi)r{\partial \tilde f \over \partial
r}(r,\varphi)\D \varphi \D r+\\
\nonumber
&+&\int_0^R{\omega(r)\over r}\sqrt{2\over\pi} \biggl( -{2k+1\over
2}\tilde f(r,0)-(-1)^k {\partial \tilde f \over \partial
\varphi}(r,\pi)+\int_0^\pi{\partial^2\tilde f \over \partial
\varphi^2}(r,\varphi)\sin{2k+1\over2}\varphi \D \varphi \biggr )\D r=\\
\nonumber
&=&\lim_{n\rightarrow\infty}\int_0^R\int_0^\pi\omega(r)
\xi_{n,k}(\varphi){\partial \over \partial r}\Bigl (r{\partial \tilde f \over \partial
r}(r,\varphi)\Bigr )\D \varphi \D r+\\
\nonumber &+&\int_0^R{\omega(r)\over
r}\sqrt{2\over\pi}\int_0^\pi{\partial^2\tilde f \over \partial
\varphi^2}(r,\varphi)\sin{2k+1\over2}\varphi \D \varphi \D r=\\
\nonumber
&=&\int_0^R\int_0^\pi\sqrt{2\over\pi}\omega(r)r\Bigl({\partial^2\tilde f
\over \partial r^2}(r,\varphi)+{1 \over r}{\partial \tilde f \over \partial
r}(r,\varphi)+{1 \over r^2}{\partial^2 \tilde f \over \partial
\varphi^2}(r,\varphi)
\Bigr )\sin{2k+1\over2}\varphi \D \varphi \D r =\\
\nonumber
&=&\int_0^R c_k(r)r\omega(r)\D r,
\end{eqnarray}
so
\begin{equation}
\label{ck}
c_k(r)=F''_k(r)+{1\over r}F'_k(r)-{1 \over
r^2}\Bigl ({2k+1 \over 2}\Bigr)^2F_k(r) {\rm \ \ for\  a.e.\ \ }r.
\end{equation}
Let us denote $\triangle_p\tilde f(r,\varphi)={\partial^2\tilde f
\over \partial r^2}(r,\varphi)+{1 \over r}{\partial \tilde f \over \partial
r}(r,\varphi)+{1 \over r^2}{\partial^2 \tilde f \over \partial
\varphi^2}(r,\varphi)$.
Taking into account (\ref{Laplace polar}) we will solve an equation
$\triangle_p\tilde f(r,\varphi)=h(r,\varphi)$ for any $h \in
L^2\bigl((0,R)\times(0,\pi),r\D r\D \varphi\bigr )$. We are seeking, of course,
only those solutions, for which the function $f(x,y)$ corresponding to
the function $\tilde f(r,\varphi)$ remains in the set $H^1\bigl (B(P,R)\cap
\Omega\bigr)$.
If we decompose a function $h$ to the series $h=\sqrt{2
\over\pi}\sum_{k=0}^\infty H_k(r)\sin{2k+1\over 2}\varphi$ we get a
set of equations:
$$
F''_k(r)+{1 \over r}F'_k(r)-{1\over r^2}\Bigl({2k+1\over
2}\Bigr)^2F_k(r)=H_k,{\rm \ \ \ }k=0,\ldots
$$
We denote $\nu=k+1/2$.
The solutions of these equations are
\begin{equation}
\label {F0}
F_0=r^{1\over2}\int_0^rH_0(z)z^{1\over2}\D z-r^{-{1\over2}}\int_0^rH_0(z)z^{3\over2}\D
z+C_1^{(0)}r^{1 \over 2}+C_2^{(0)}r^{-{1\over2}}
\end{equation}
and for $k>0$
\begin{equation}
\label{Fk}
F_k={1\over 2\nu}r^\nu\int_R^rH_k(z)z^{-\nu+1}\D z-{1\over 2\nu}r^{-\nu}\int_0^rH_k(z)z^{\nu+1}\D
z+C_1^{(k)}r^\nu+C_2^{(k)}r^{-\nu}.
\end{equation}
We can compute the first derivatives
\begin{equation}
\label {F'0}
F'_0={1\over 2}r^{-{1\over2}}\int_0^rH_0(z)z^{1\over2}\D z+{1\over 2}
r^{-{3\over2}}\int_0^rH_0(z)z^{3\over2}\D
z+{1\over2}C_1^{(0)}r^{-{1 \over 2}}-{1\over2}C_2^{(0)}r^{-{3\over2}}
\end{equation}
and for $k>0$
\begin{equation}
F'_k={1\over 2}r^{\nu-1}\int_R^rH_k(z)z^{-\nu+1}\D z+{1\over 2}r^{-\nu-1}\int_0^rH_k(z)z^{\nu+1}\D
z+\nu C_1^{(k)}r^{\nu-1}-\nu C_2^{(k)}r^{-\nu-1}.
\label{F'k}
\end{equation}
Because $\tilde f$, ${\partial \tilde f \over \partial r}\in
L^2\bigl((0,R)\times(0,\pi),r\D\varphi\D r\bigr)$, the functions
$F_k$ and $F'_k$ have to be in the set $L^2\bigl ((0,R),r \D
r\bigr)$ for all $k$.
Taking the first two terms in (\ref{Fk}) we get after application
of the triangle and Schwarz inequalities
\begin{eqnarray}
\nonumber
&{\rm \ }&\Bigl | {1\over 2\nu}r^\nu\int_R^rH_k(z)z^{-\nu+1}\D z-
{1\over 2\nu}r^{-\nu}\int_0^rH_k(z)z^{\nu+1}\D z \Bigr |\leq\\
\nonumber
&\leq&{1\over2\nu}\Bigl(r^\nu\sqrt{\int_r^R|H_k(z)|^2z\D z}
\sqrt{\int_r^Rz^{-2\nu+1}\D z}+r^{-\nu}\sqrt{\int_0^r|H_k(z)|^2z\D z}
\sqrt{\int_0^rz^{2\nu+1}\D z}\Bigr)\leq\\
\nonumber
&\leq&{1\over2\nu}r\sqrt{\int_0^R|H_k(z)|^2z\D z}\Bigl
(\sqrt{1-\bigl({R\over r}\bigr)^{-2\nu+2}\over 2\nu-2}+{1\over
\sqrt{2\nu+2}}\Bigr)\leq\\
\nonumber
&\leq&{r\over \nu\sqrt{2\nu-2}}\sqrt{\int_0^R|H_k(z)|^2z\D z}
\end{eqnarray}
for $0<r<R$.
Using similar procedure on (\ref{F0}), (\ref{F'0}) and (\ref{F'k})
we get the following inequalities holding for every $k=0,1,\ldots$
\begin{eqnarray}
\label{odhad Fk}
\bigl |F_k-C_1^{(k)}r^\nu-C_2^{(k)}r^{-\nu} \bigr| \leq
{r\over \nu\sqrt{|2\nu-2|}}\sqrt{\int_0^R|H_k(z)|^2z\D z}\\
\nonumber\\
\label{odhad F'k}
\bigl |F'_k-\nu C_1^{(k)}r^{\nu-1}+\nu C_2^{(k)}r^{-\nu-1} \bigr| \leq
{1\over \sqrt{|2\nu-2|}}\sqrt{\int_0^R|H_k(z)|^2z\D z}.
\end{eqnarray}
We conclude using this estimate that the first three terms in
(\ref{Fk}) belong to $L^2\bigl ((0,R),r \D r\bigr)$ and thus the
forth term has to be in this set too.
But it is obvious that $r^{-\nu} \not\in L^2\bigl ((0,R),r \D
r\bigr)$ for $\nu\geq3/2$.
Hence $C_2^{(k)}=0$ for $k\geq1$.
Applying the same arguments to (\ref{F'0}) we have also
$C_2^{(0)}=0$.
Moreover the condition $\int_0^R\sum_{k=0}^\infty|F'_k(r)|^2r \D r
<\infty$ must be satisfied.
First we suppose $C_1^{(k)}=0$ for all $k$.
Then using estimate (\ref{odhad F'k})
\begin{eqnarray}
\nonumber
\int_0^R\sum_{k=0}^\infty|F'_k(r)|^2r \D r &\leq& \int_0^R r \D r
\sum_{k=0}^\infty {1\over 2|\nu-1|}\int_0^R|H_k(z)|^2z\D z\leq\\
\nonumber
&\leq& \|h\|^2_{L^2\bigl((0,R)\times(0,\pi),r\D\varphi\D
r\bigr)}\int_0^Rr\D r=\\
\nonumber
&=&{R^2\over 2}\|h\|^2_{L^2\bigl((0,R)\times(0,\pi),r\D\varphi\D
r\bigr)}.
\end{eqnarray}
Thus we must choose the constants $C_1^{(k)}$ so as
\begin{equation}
\label{odhad C1}
\sum_{k=0}^\infty\int_0^R|C_1^{(k)}|^2\nu^2r^{2\nu-2}r\D r=
\sum_{k=0}^\infty |C_1^{(k)}|^2{\nu \over 2}R^{2\nu}<\infty.
\end{equation}
Now we turn our attention to the function $\tilde g(r,\varphi)$.
Using (\ref{H1 polar}) we can write
$$
u(r):=\sum_{k=0}^\infty{1\over
r^2}\Bigl({2k+1\over2}\Bigr)^2\bigl|G_k(r)\bigr|^2 \in
L^1\bigl((0,R),r\D r\bigr).
$$
Denoting $v(r)=r u(r)$ we get a function $v\in
L^1(0,R)$.
Thus we can state $\liminf_{r\rightarrow 0^+} \bigl|r v(r)\bigr|=0$.
Otherwise $\bigl |r v(r)\bigr|\geq a>0$ for a.e. small $r$, so
$|v(r)|\geq a/r$ for some constant $a$ and a.e. sufficiently small
$r$, which is a contradiction with the fact that $v\in L^1(0,R)$.
Hence we can find a sequence $\{r_n\}_{n=1}^\infty$ such that
$r_n\rightarrow 0^+$ and
\begin{equation}
\label{odhad G}
\lim_{n\rightarrow\infty}\bigl |r_n
v(r_n)\bigr|=\lim_{n\rightarrow\infty}\sum_{k=0}^\infty
\Bigl({2k+1\over2}\Bigr)^2\bigl|G_k(r_n)\bigr|^2=0.
\end{equation}
Now we are ready to return to (\ref{povrchovy clen}).
We rewrite the particular terms of this sum using (\ref{rozklad
f}) and (\ref{rozklad g}).
$$
\int_0^\pi {\partial\tilde f_k(r_n,\varphi)
\over \partial r}\overline{\tilde g_k(r_n,\varphi)} r_n \D \varphi
= \sum_{k=0}^\infty r_n F'_k(r_n)\overline{G_k(r_n)}
$$
The r.h.s. of this equality can be estimated using (\ref{odhad
F'k}), the triangle inequality and the Schwarz
inequality in the space $\ell^2$.
\begin{eqnarray}
\nonumber
\Bigl| \sum_{k=0}^\infty r_n F'_k(r_n)\overline{G_k(r_n)}\Bigr|
&\leq&r_n \sum_{k=0}^\infty \bigl |G_k(r_n)\bigr|
\Bigl(\sqrt{\int_0^R\bigl|H_k(z)\bigr|^2 z\D z}{1\over
\sqrt{|2\nu-2|}}+\nu\bigl|C_1^{(k)}\bigr|r_n^{\nu-1}\Bigr)\leq\\
\nonumber
&\leq&  \sqrt{\sum_{k=0}^\infty\bigl|G_k(r_n)\bigr|^2}\Biggl (r_n
\|h\|_{L^2\bigl((0,R)\times(0,\pi),r\D\varphi\D r\bigr)}+
\sqrt{\sum_{k=0}^\infty\nu^2\bigl|C_1^{(k)}\bigr|^2r_n^{2\nu}}\Biggr)\\
\label{odhad povrchoveho clenu}
\end{eqnarray}
Because of (\ref{odhad G}) it is enough to show, that the last
term in brackets in (\ref{odhad povrchoveho clenu}) is bounded and we will
know that the statement (\ref{povrchovy clen}) holds.
Due to (\ref{odhad C1}) it is sufficient to prove that the inequality
$\nu r^{2\nu}<R^{2\nu}/2$ holds for all $\nu=1/2,3/2,\ldots$ and
sufficiently small $r$.
Rewriting the inequality to the form
\begin{equation}
\label{r/R}
{r\over R}<\Bigl({1\over 2\nu}\Bigr)^{1\over 2\nu}
\end{equation}
we will study the function $\phi(x)=\bigl({1\over x}\bigr)^{1\over x}$ on
the interval $[1,\infty)$.
It is obvious that $\phi$ is continuous and strictly positive on this interval.
As
$$
{d\over d x}\phi(x)=\phi(x){\ln(x)-1\over x^2}
$$
$\phi$ reaches its global minimum at the point $x=e$.
Thus for $r/R<(1/e)^{1/e}$ the inequality (\ref{r/R}) holds for
every $\nu$, which completes the proof.
\qed
\end{PF}
\begin{rem}
The fact, that the all functions from the domain of the Hamiltonian,
which is defined by the quadratic form, satisfy the right
boundary condition is supposed to be well-known for domains with reasonable
boundaries.
But it is also very often supposed, that the operator domain is
the subset of the Sobolev space $H^2(\Omega)$, which satisfies the
boundary conditions.
The systems with combined boundary conditions are examples of operators, for
which this assertion is not true.
The situation is similar to the systems studied in \cite{BS}.
It was shown there, that for bounded regions in the plane
$\mathbb{R}^2$ with piecewise $C^3$-boundary, which has finite
number of angles larger than $\pi$, there exists for each such
angle one function, which is not in $H^2(\Omega)$ and which belongs to
the operator domain with the Dirichlet boundary condition.
The operator domain is then a span of the $H^2(\Omega)$ space and all these functions, called
Guseva functions.

It is easy to check, that similar functions belong to operator
domain in our systems too.
Let us take the following function, written in polar coordinates with the
origin in some $P_k$ and $R<d$.
\begin{equation}
\label{Guseva}
f(r,\varphi)=\xi(r)r^{1\over 2}\sin{\varphi\over2},
\end{equation}
where $\xi(r)\in C^\infty(0,\infty)$, $\xi(r)=1$ on $(0,R/2)$ and
$\xi(r)=0$ for $r>R$.
This function satisfies the boundary conditions, $\triangle_pf \in
L^2(\Omega)$ and $f\in H^1(\Omega)$, thus $f\in D(\Omega)$.
But $f\not\in H^2(\Omega)$.
The trace on the part of the boundary $(-R/2,R/2)$ is
$$
{1\over r}{\partial f \over \partial \varphi}(r,\varphi)=\left \{
\begin{array}{lcl}
{1\over 2}r^{-1/2} & {\rm for} & \varphi=0\\
\\
0 & {\rm for} & \varphi=\pi.
\end{array} \right .
$$
So even the trace of the normal derivatives on the boundary is not square
integrable.
These are the reasons why we cannot immediately use the Gauss
Theorem in the proof of Theorem 1.

The interesting open question arises whether all eigenfunctions
have Guseva-like behaviour near the points $P_k$.
\end{rem}

\section{Bound states}
\setcounter{equation}{0}

Now we are going to study our specific systems from Fig.~1.
First we localize the essential spectra of these systems and make
the first estimate on the number of bound states below the
essential spectrum treshold using the technique of the
Dirichlet-Neumann bracketing (see \eg\ \cite{RS}, Section XIII.15).
Then we will continue with the specification of the number of
bound states using variational methods.

\subsection{Essential spectrum, number of bound states}

Following arguments are the same for both our system, so we do not
distinguish between them in this subsection.
Cutting the domain $\Omega$ by the additional Neumann or Dirichlet
boundaries parallel to the $y-$axis at $x=\pm\delta$, we get new
operators $H^{(N)}$, $H^{(D)}$ defined in the standard way, using
the quadratic form.
We can decompose these operators $H^{(j)}=H^{(j)}_t\oplus
H^{(j)}_c$, $j=N,D$, where the ``tail" part corresponds to the two
halfstrips and the rest to the central part with the Neumann and
Dirichlet condition on the vertical boundaries, respectively.
Using Dirichlet-Neumann bracketing we have $H^{(N)}_t\oplus
H^{(N)}_c \leq -\triangle_{DN}^\Omega\leq H^{(D)}_t\oplus
H^{(D)}_c$
in the sense of quadratic forms (see \cite{RS}, Section XIII.15).

Now $\sigma_{ess}\bigl(H^{(j)}_t\bigr)=[{\pi^2\over 4
d^2},\infty)$, $j=N,D$ (we get this result after simple
calculation using Example 4.9.6 in \cite{BEH} and Corollary of the
Theorem VIII.33 in \cite{RS}).
By the minimax principle (see \eg\ \cite{RS}, Section XIII.1)
$-\triangle_{DN}^\Omega$ has the same infimum of the essential
spectrum.
To verify that $\sigma_{ess}\bigl(-\triangle_{DN}^\Omega\bigr)$ is
indeed the whole interval $[{\pi^2\over 4 d^2},\infty)$ we can use
the same procedure as for $H_t^{(j)}$ (see Example 4.9.6 in
\cite{BEH}).
Possible isolated eigenvalues of
$-\triangle_{DN}^\Omega$ are squeezed between those of
$H_c^{(j)}$, $j=N,D$.
Because the first eigenvalue of $H_c^{(N)}$ is zero,
$-\triangle_{DN}^\Omega$ has an eigenvalue below the essential
spectrum treshold provided $H_c^{(D)}$ does, which is true if
$\delta> d$.

More generally, the number $N_D$ of eigenvalues of $H_c^{(D)}$
smaller than ${\pi^2 \over 4 d^2}$ equals the largest integer
number smaller than ${\delta \over d}$, \ie\ $N_D=-\bigl [-\delta/d\bigr]-1$,
where $[\cdot]$ denotes the entire part.
The number of the ``Neumann" eigenvalues of $H_c^{(N)}$ is $N_N=1+N_D$.
This means that the number of bound states of
$-\triangle_{DN}^\Omega$ below the essential spectrum treshold satisfies the
inequality
\begin{equation}
\label{pocet vlastnich hodnot}
-\Bigl[-{\delta \over d}\Bigr ]-1\leq N \leq -\Bigl[-{\delta \over d}\Bigr
].
\end{equation}
We see that $-\triangle_{DN}^\Omega$ has isolated
eigenvalues, at least for $\delta$ large enough.
In the same way, one finds that the $m$--th eigenvalue $\mu_m$ of $-\triangle_{DN}^\Omega$
is estimated by
\begin{equation}
\label{odhad vlastni hodnoty}
\Bigl({m-1 \over \lambda}\Bigr)^2 \leq {\mu_m \over \mu} \leq \Bigl({m \over
\lambda}\Bigr)^2,
\end{equation}
where $\lambda=\delta/d$ and $\mu=\inf(\sigma_{ess})={\pi^2\over
4 d^2}$, and that the critical value $\lambda_m=\delta_m/d$ at
which $m$--th eigenvalue appears satisfies the bounds
\begin{equation}
\label{kriticka lambda}
m-1 \leq \lambda_m \leq m.
\end{equation}
To learn more about the dependence of the eigenvalues and the
corresponding eigenfunctions on $\lambda$, we have to use a
different technique.

\subsection{Existence of bound states}

The above existence argument for $\lambda > 1$ is a crude
one.
In fact, there is no lower bound on the length of the overlay of
Neumann boundaries for case B).
On the other hand we will show that in system A) the discrete
spectrum of the Hamiltonian is empty for small $\lambda$, but the
ground state appears sooner than $\lambda=1$.
We will distinguish our two cases writing
$-\triangle_{DN}^{\Omega,A}$, resp. $-\triangle_{DN}^{\Omega,B}$
instead of $-\triangle_{DN}^\Omega$ and $Q^A(q_0)$, resp.
$Q^B(q_0)$ instead of $Q(q_0)$.
\begin{thm}
\label{groundstate B}
The operator $-\triangle_{DN}^{\Omega,B}$ has an isolated eigenvalue
in $[0,\mu)$ for any $\delta>0$.
\end{thm}
\begin{PF}
We slightly modify for the present purpose the variational
proof of the Theorem in \cite{ESTV}, which comes out from the variational
argument of \cite{GJ}. The transverse ground-state wavefunction at
the ``tails" of our strip is $$\chi(y)=\sqrt{2\over
d}\cos\sqrt\mu y.$$ For any $\Phi \in Q^B(q_0)$ we put
\begin{equation}
\label{q}
q[\Phi]=q_0(\Phi,\Phi)-\mu\|\Phi\|^2_{L^2(\Omega)}.
\end{equation}

Since the essential spectrum of $-\triangle_{DN}^{\Omega,B}$
starts at $\mu$, we have to find a trial function $\Phi$ such
that $q[\Phi]<0$, it has to belong to the form domain $Q^B(q_0)$
(see \eg\ \cite{D}, Chapter 4).
Thus in particular we can choose $\Phi$ continuous inside $\Omega$, but not
necessarily smooth.
Notice first that if
$\,\Phi(x,y)= \varphi(x) \chi(y)\,$, we have
   \begin{equation} \label{q form}
q[\Phi]\,=\, \|\varphi'\|^2_{L^2(\mathbb{R})}\,.
   \end{equation}
To make the longitudinal contribution to the kinetic energy
small, we use an external scaling. We choose an interval
$\,J=[-b,b]\,$ for a positive $\,b>\delta\,$ and a function
$\,\varphi\in\mathcal{S}(\mathbb{R})\,$ such that $\,\varphi(x)=1\,$ if
$\,x\in J\;$; then we define the family $\,\{ \varphi_{\sigma}
| \sigma>0\,\}\,$ by
\begin{equation} \label{phi sigma}
\varphi_{\sigma}(x)\,=\, \left\lbrace\: \begin{array}{lll}
\varphi(x) & {\rm for} & |x|\le b \\ \\
\varphi(\pm b+\sigma(x\mp b)) & {\rm for} & |x|\ge b    \end{array}
\right.
\end{equation}
Finally, let us choose a real localization function $\,j\in
C_0^{\infty}(-\delta,\delta)\,$ and define
   \begin{equation} \label{trial B}
\Phi_{\sigma,\eps}(x,y)= \varphi_{\sigma}(x)
\bigl (\chi(y)+\eps j(x)^2 \bigr )
   \end{equation}
for any $\sigma,\,\eps>0$.
The main point of the construction is that we modify the
factorized function we started with in two mutually disjoint
regions, outside and inside the rectangle $\,J\times
(0,d)\,$. Hence the functions $\,\varphi'_{\sigma}\,$
and $\,j^2\,$ have disjoint supports. Using this together
with the identity
\begin{equation}\label{phi'sigma}
\|\varphi'_{\sigma}\|^2_{L^2(\mathbb{R})}=
\sigma\|\varphi'\|^2_{L^2(\mathbb{R})},
\end{equation}
the explicit form of the function $\chi$ and (\ref{q form}),
we get after straightforward calculation
\begin{equation}
\label{trial value}
q[\Phi_{\sigma,\eps}]=\sigma \|\varphi'\|^2_{L^2(\mathbb{R})} -
\eps{\pi \over d}\sqrt{2\over d}\, \|j\|^2_{L^2(\mathbb{R})} + \eps^2 \bigl(
4d\|j j'\|^2_{L^2(\mathbb{R})}-d\mu\|j^2\|^2_{L^2(\mathbb{R})} \bigr).
\end{equation}
By construction, the last two terms on the right hand side of
(\ref{trial value}) are independent of $\sigma$.
Moreover, the term linear in $\eps$ is negative, so choosing $\eps$
sufficiently small, we can make it dominate over the
quadratic one. Finally, we fix this $\eps$ and
choose a small enough $\sigma$ to make the right hand
side of (\ref{trial value}) negative. \qed
\end{PF}

Now we move to the case A),where the situation is more complicated.
\begin{thm} There exists a real number $\Lambda_0\in (0,1)$, such that the
discrete spectrum of the operator
$-\triangle_{DN}^{\Omega,A}(\lambda)$ is empty for all $\lambda
\leq \Lambda_0$ and there exists at least one isolated eigenvalue
in the spectrum of this operator for all $\lambda > \Lambda_0$.
\end{thm}
\begin{PF}
Taking into account that $\mu_1(\lambda)=\inf_{\varphi \in Q^A(q_0),\|\varphi\|=1}
q_0(\varphi)$ is a nonincreasing continuous function of $\lambda$
(see Appendix A) and the minimax principle \cite{RS}, it is
sufficient to show that there are two real positive numbers
$\Lambda_1<\Lambda_2$, $\Lambda_1, \Lambda_2\in (0,1)$, such that
\begin{itemize}
\item[\textrm{(i)}]
the discrete spectrum is empty for all $\lambda<\Lambda_1$,
\item[\textrm{(ii)}]
there exists at least one isolated eigenvalue in the spectrum for
all $\lambda>\Lambda_2$.
\end{itemize}

We know from the previous subsection that there exists a bound state
for $\lambda>1$.
Let us search for better estimate $\Lambda_2<1$ by the similar variational
technique as in the proof of the Theorem 2.
We are seeking the trial function $\Phi\in Q^A(q_0)$, for which
the functional $q$ defined by (\ref{q}) has a negative value.
We again choose $\Phi$ continuous inside $\Omega$, but not
necessarily smooth and we use the same trick to make the
longitudinal contribution to the kinetic energy small.
The problem is, that since we have different transverse
ground-state wavefunctions at both tails (region I, resp. III on
the Fig.1), we cannot construct the ``support function" (like
$\varphi_\sigma\chi$ in the proof of Theorem 2), which
longitudinal derivative has disjoint support with the localization
function (like $\eps j$ before).

We start with the trial function
\begin{equation} \label{trial A}
\Phi_\sigma(x,y)=\varphi_\sigma(x)\sin{\pi y\over
2d}+\psi_\sigma(x)\cos{\pi y \over 2d}+\eta(x)+\chi(x)\cos{\pi
y\over d},
\end{equation}
where, similarly to (\ref{phi sigma})
\begin{eqnarray}
\nonumber
\varphi_\sigma(x)&=&\left \{\begin{array}{lcl}
\varphi(x) & {\rm for} & x\geq -\delta\\
\\
\varphi\bigl(-\delta+\sigma(x+\delta)\bigr) & {\rm for} & x\leq
-\delta
\end{array} \right .
\nonumber\\
\nonumber\\
\nonumber
\psi_\sigma(x)&=&\left \{\begin{array}{lcl}
\psi(x) & {\rm for} & x\leq \delta\\
\\
\psi\bigl(\delta+\sigma(x-\delta)\bigr) & {\rm for} & x\geq
\delta,
\end{array} \right .
\end{eqnarray}
$\varphi$, $\psi\in \mathcal{S}(\mathbb{R})$,
$\varphi(-\delta)=\psi(\delta)=1$; $\varphi(x)=0$ on
$[\delta,\infty)$, $\psi(x)=0$ on $(-\infty,-\delta]$, $\eta$,
$\chi \in C^\infty[-\delta,\delta]$ and $\eta(x) =\chi(x)=0$ for
$|x|\geq\delta$.
The value 1 of $\varphi(-\delta)$ and $\psi(\delta)$ is not
important, it can be any constant without the influence on the
result.
We choose $\varphi(-\delta)=\psi(\delta)$, because we expect the
ground state to be symmetric.
We shall assume the functions $\varphi$, $\psi$, $\eta$, $\chi$ to
be real.
Here we can compare this trial function with that one from
(\ref{trial B}).
The role of $\varphi_\sigma$ from previous case plays here the
functions $\varphi_\sigma$ and $\psi_\sigma$, while the role of localization
function plays here $\eta+\chi\cos{\pi y \over d}$.
We decompose the functional $q$ to three parts, in each of which
one integrates over the region I, resp.II, resp.III
$$
q[\Phi_\sigma]=q_I[\Phi_\sigma]+q_{II}[\Phi_\sigma]+q_{III}[\Phi_\sigma]
={d \over 2}\,\sigma\bigl(\|\varphi'\|^2_{L^2(-\infty,-\delta)}+
\|\psi'\|^2_{L^2(\delta,\infty)}\bigr )+q_{II}[\Phi_\sigma],
$$
where we used identities similar to (\ref{phi'sigma}).
We see that the first term here is always positive, but it
can be arbitrarily small, due to parameter $\sigma$, while
the second term does not depend on $\sigma$.
We easily compute
\begin{eqnarray}\nonumber
q_{II}[\Phi_\sigma]&=&\int_{-\delta}^\delta\biggl
({d\over
2}\,\bigl(\varphi'(x)^2+\psi'(x)^2+\chi'(x)^2\bigr)+d\eta'(x)^2 +{2d
\over \pi}\, \varphi'(x)\psi'(x)+\\
\nonumber &+&{4d\over 3\pi}\,\chi'(x)\bigl(\psi'(x)-\varphi'(x)
\bigr)+{4d \over \pi}\,\eta'(x)\bigl(\varphi'(x)+\psi'(x)\bigr) +
{\pi \over d}\,\chi(x)\bigl(\psi(x)-\varphi(x)\bigr)+\\
\label{qII} &+& {3\pi^2 \over 8d}\,\chi(x)^2-{\pi\over d}\,\varphi(x)
\psi(x)-{\pi^2\over 4d}\,\eta(x)^2-{\pi \over d}\,\eta(x)\bigl(\psi(x)
+\varphi(x)\bigr)\biggr)\D x.
\end{eqnarray}
We choose the solution of the Euler equations
\begin{eqnarray}\nonumber
d\varphi''(x)+{2d\over \pi}\,\psi''(x)-{4d\over 3\pi}\,\chi''(x)
+{4d \over \pi}\,\eta''(x)+{\pi\over
d}\,\bigl(\psi(x)+\eta(x)+\chi(x)\bigr)&=&0\\
\nonumber
d\psi''(x)+{2d\over \pi}\,\varphi''(x)+{4d\over 3\pi}\,\chi''(x)
+{4d \over \pi}\,\eta''(x)+{\pi\over
d}\,\bigl(\varphi(x)+\eta(x)-\chi(x)\bigr)&=&0\\
\nonumber
d\chi''(x)+{4d\over 3\pi}\,\bigl(\psi''(x)-\varphi''(x)\bigr)
-{\pi\over d}\,\bigl(\psi(x)-\varphi(x)\bigr)-{3\pi^2\over
4d}\,\chi(x)&=&0\\
\nonumber
2d\eta''(x)+{4d\over \pi}\,\bigl(\psi''(x)+\varphi''(x)\bigr)
+{\pi\over d}\,\bigl(\psi(x)+\varphi(x)\bigr)+{\pi^2\over
2d}\,\eta(x) &=&0
\end{eqnarray}
to be a trial function.
By linear combinations of these equations we can obtain uncoupled
second order differential equations for $\varphi-\psi$ and
$\varphi+\psi$.
Then the solution with boundary condition mentioned above is
obtained,
\begin{eqnarray}\nonumber
\chi(x)&=&{4\over 3\pi}\,\Biggl({\sinh{\sqrt3\pi x\over 2d}\over
\sinh{\sqrt3\pi \delta\over2d}}-{\sinh{\pi x\over d}\sqrt{3{3\pi-8\over
9\pi^2-18\pi-32}}\over {\sinh{\pi \delta\over d}\sqrt{3{3\pi-8\over
9\pi^2-18\pi-32}}}}\Biggr )\\
\nonumber\\
\nonumber\\
\nonumber
\varphi(x)&=&{1\over 2}\,\Biggl({\cosh{\pi x\over d}\sqrt{4-\pi\over\pi^2+2\pi-16}\over
\cosh{\pi \delta\over d}\sqrt{4-\pi\over\pi^2+2\pi-16}}-{\sinh{\pi x\over d}\sqrt{3{3\pi-8\over
9\pi^2-18\pi-32}}\over {\sinh{\pi \delta\over d}\sqrt{3{3\pi-8\over
9\pi^2-18\pi-32}}}}\Biggr )\\
\label{vysledek Euler}\\
\nonumber
\psi(x)&=&{1\over 2}\,\Biggl({\cosh{\pi x\over d}\sqrt{4-\pi\over\pi^2+2\pi-16}\over
\cosh{\pi \delta\over d}\sqrt{4-\pi\over\pi^2+2\pi-16}}+{\sinh{\pi x\over d}\sqrt{3{3\pi-8\over
9\pi^2-18\pi-32}}\over {\sinh{\pi \delta\over d}\sqrt{3{3\pi-8\over
9\pi^2-18\pi-32}}}}\Biggr )\\
\nonumber\\
\nonumber\\
\nonumber
\eta(x)&=&{2\over\pi}\,\Biggl({\cos{\pi x\over 2d}\over
\cos{\pi \delta\over 2d}}-{\cosh{\pi x\over d}\sqrt{4-\pi\over\pi^2+2\pi-16}\over
\cosh{\pi \delta\over d}\sqrt{4-\pi\over\pi^2+2\pi-16}}\Biggr ).
\end{eqnarray}
As the quadratic form of the derivatives in (\ref{qII}) is positive
definite this trial function is a good candidate for a minimum of the
functional (\ref{qII}).
Now we substitute (\ref{vysledek Euler}) to (\ref{qII}) and
after a tedious but straightforward calculation we obtain
\begin{eqnarray} \nonumber
q_{II}[\Phi_\sigma]&=&
{\sqrt{(4-\pi)(\pi^2+2\pi-16)}\over 2\pi}\,\tanh
{\pi \delta\over d}\sqrt{4-\pi\over\pi^2+2\pi-16} + {8\over
3\sqrt3\pi}\,\coth{\sqrt3\pi\delta\over2d}+\\
\label{value qII}\\
\nonumber
&+&{\sqrt{(3\pi-8)(9\pi^2-18\pi-32)}\over 6\sqrt3\pi} \,\coth{\pi \delta\over d}\sqrt{3{3\pi-8\over
9\pi^2-18\pi-32}}-{4\over \pi}\,\tan{\pi\delta\over2d}.
\end{eqnarray}
Now we can understand $q_{II}$ like a function of a variable
$\delta$.
We see, that $q_{II}$ is a continuous function on the interval
$(0,d)$, $\lim_{\delta \rightarrow 0^+}q_{II}(\delta)=+\infty$ and
$\lim_{\delta\rightarrow d^-}q_{II}(\delta)=-\infty$.
So there must exist a point $\delta_0\in (0,d)$ and corresponding number
$\Lambda_2=\delta_0/d$, such that $q_{II}(\delta)<0$ for
$\delta\in(\delta_0,d)$.
Thus we can find for every $\delta$ from this interval a number
$\sigma$ small enough to have $q[\Phi_\sigma]<0$, which finishes
the proof of the existence of $\Lambda_2$.

Now we are going to prove, that the discrete spectrum of the
operator $-\triangle^{\Omega,A}_{DN}(\lambda)$ is empty for all
$\lambda\leq \Lambda_1$.
It will be shown if we demonstrate that the functional
$q[\Phi]\geq 0$ for all $\Phi$ from a suitable dense (in
$H^1(\Omega)$ norm) set in $Q^A(q_0)$, say $$\mathcal{Q}(\Omega)=\bigl \{\psi \in
H^1(\Omega)\cap
C(\overline\Omega)\,|\,\psi \in C^2(\Omega_j),\,j=I,II,III,\,\psi\upharpoonright\mathcal{D}=0\bigr\},$$
where $C(\overline\Omega)$ is just the set of functions continuous
in the closure of $\Omega$.
It can be proven that this set is really dense in $Q^A(q_0)$ (See Appendix
B).
We again decompose $q$ into three parts
$q[\Phi]=q_I[\Phi]+q_{II}[\Phi]+q_{III}[\Phi].$
The ``tail" parts of $\Phi \in \mathcal{Q}(\Omega)$ we expand to the series
\begin{equation}\label{PhiI}
\Phi(x,y)=\sqrt{2\over d}\,\sum_{k=0}^\infty a_k(x)\sin{2k+1\over
2d}\pi y
\end{equation}
in the region $\Omega_I$ and
\begin{equation}\label{PhiIII}
\Phi(x,y)=\sqrt{2\over d}\,\sum_{k=0}^\infty b_k(x)\cos{2k+1\over
2d}\pi y
\end{equation}
in the region $\Omega_{III}$.
Using the same procedure like in the proof of the Theorem 1, we
know that these series can be differentiated by terms.
Hence
\begin{eqnarray}
\label{qI}
q_I[\Phi]&=& \sum_{k=0}^\infty \,\int_{-\infty}^{-\delta}\Bigl
(\bigl|a'_k(x)\bigr|^2 + {\pi^2\over d^2}\,(k^2+k)\bigl
|a_k(x)\bigr |^2\Bigr)\D x\\
\nonumber\\
\label{qIII}
q_{III}[\Phi]&=&\sum_{k=0}^\infty \, \int_\delta^\infty\Bigl
(\bigl|b'_k(x)\bigr|^2 + {\pi^2\over d^2}\,(k^2+k)\bigl
|b_k(x)\bigr |^2\Bigr)\D x
\end{eqnarray}
and we see that $q_I[\Phi]\geq 0$ and $q_{III}[\Phi]\geq 0$ for
all $\Phi \in \mathcal{Q}(\Omega)$.
As we have seen above (similarly to (\ref{phi'sigma})) the contributions from the
terms
$|a'_0(x)|^2$ and $|b'_0(x)|^2$ can be arbitrarily small.
We minimalize the rest of the functional using Euler equations, the
general boundary condition fixing the function values at $x=\delta$,
resp. $x=-\delta$, and the square integrability.
As $\Phi \in C(\overline\Omega)$ the Fourier coefficients $a_k$,
resp. $b_k$, are continuous in $(-\infty,-\delta]$, resp.
$[\delta,\infty)$, and $a_k(-\delta)$, resp. $b_k(-\delta)$, are
those of $\Phi(-\delta,\cdot)$, resp. $\Phi(\delta,\cdot)$.
The solution of the Euler equations will be really the absolute
minimum of $q_I+q_{III}$.
It is easy to see, that for all $\Phi$ in  the set $\mathcal{Q}(\Omega)$,
$q_I[\Phi]+q_{III}[\Phi]=q_I[\Phi_0]+q_{III}[\Phi_0]+
q_I[\Phi-\Phi_0]+q_{III}[\Phi-\Phi_0]$, where $\Phi_0$ is given by
Euler equations and we have seen above that
$q_I[\Phi-\Phi_0]+q_{III}[\Phi-\Phi_0]\geq 0$.
For $k=1,\ldots,\infty$ we get
\begin{eqnarray}\nonumber
a''_k(x)-{\pi^2\over d^2}\,k(k+1)a_k(x)&=&0\\
\nonumber\\
\nonumber
b''_k(x)-{\pi^2\over d^2}\,k(k+1)b_k(x)&=&0
\end{eqnarray}
and the square integrable solutions of these equations are
\begin{eqnarray}\label{Ak}
a_k(x)&=&A_k\exp\bigl({\pi\over d}\,\sqrt{k(k+1)}(x+\delta)\bigr )\\
\nonumber\\
\label{Bk}
b_k(x)&=&B_k\exp\bigl({\pi\over d}\,\sqrt{k(k+1)}(\delta-x)\bigr
).
\end{eqnarray}
Putting these results to (\ref{qI}) and (\ref{qIII}) we have
\begin{equation}
\label{qI+qIII}
q_I[\Phi]+q_{III}[\Phi] \geq {\pi\over
d}\,\sum_{k=1}^\infty\sqrt{k(k+1)}(|A_k|^2+|B_k|^2).
\end{equation}

Let us turn our attention to the functional $q_{II}$.
First we estimate the value $\|\Phi\|^2_{L^2(\Omega_{II})}$.
Denoting $\Phi_-(y)=\Phi(-\delta,y)$ and
$\Phi_+(y)=\Phi(\delta,y)$ we can write
\begin{eqnarray}\nonumber
\|\Phi\|^2_{L^2(\Omega_{II})}&=&\int_0^d\int_{-\delta}^\delta\bigl
|\Phi_-(y)+\int_{-\delta}^x{\partial \Phi \over \partial x}\,
(z,y)\D z\bigr |\,\,\bigl
|\int_{x}^\delta{\partial \Phi \over \partial x}\,
(z,y)-\Phi_+(y)\D z\bigr |\D x\D y\leq\\
\nonumber\\
\nonumber
&\leq&\int_0^d\int_{-\delta}^\delta\biggl ( |\Phi_-(y)|+\int_{-\delta}^\delta\Bigl
|{\partial \Phi \over \partial x}\,
(z,y)\Bigr |\D z\biggr )\biggl( |\Phi_+(y)|+\int_{-\delta}^\delta\Bigl
|{\partial \Phi \over \partial x}\,
(z,y)\Bigr |\D z\biggr )\D x\D y=\\
\nonumber\\
\nonumber
&=&2\delta\int_0^d\Biggl(|\Phi_-(y)|\,\,|\Phi_+(y)|+\Bigl
(|\Phi_-(y)|+ |\Phi_+(y)|\Bigr )\int_{-\delta}^\delta\Bigl
|{\partial \Phi \over \partial x}\,(z,y)\Bigr |\D z+\\
\nonumber\\
\nonumber
&+&\biggl (\int_{-\delta}^\delta\Bigl
|{\partial \Phi \over \partial x}\,(z,y)\Bigr |\D z \biggr )^2
\Biggr) \D y\leq 2\delta \Biggl ( \|\Phi_-\|_{L^2(0,d)}\,
\|\Phi_+\|_{L^2(0,d)}+\\
\nonumber\\
\nonumber
&+&\bigl{\|}\,|\Phi_-|+|\Phi_+|\,\bigr {\|}_{L^2(0,d)}\sqrt{\int_0^d
\biggl ( \int_{-\delta}^\delta\Bigl
|{\partial \Phi \over \partial x}\,(z,y)\Bigr |\D z \biggr )^2 \D
y}+ 2\delta \Bigl{\|}{\partial \Phi \over \partial
x}\Bigr{\|}^2_{L^2(\Omega_{II})} \Biggr ) \leq\\
\nonumber\\
\nonumber
&\leq&2\delta\biggl ( \|\Phi_-\|_{L^2(0,d)}\,
\|\Phi_+\|_{L^2(0,d)}+\sqrt{2\delta}\bigl{\|}\,|\Phi_-|+|\Phi_+|\,\bigr
{\|}_{L^2(0,d)} \Bigl{\|}{\partial \Phi \over \partial
x}\Bigr{\|}_{L^2(\Omega_{II})}+\\
\nonumber
&+&2\delta\Bigl{\|}{\partial \Phi \over \partial
x}\Bigr{\|}^2_{L^2(\Omega_{II})} \biggr ),
\end{eqnarray}
where we used the fact that $\Phi\in C^2(\Omega_{II})$ and the Schwarz inequality.
Denoting $\kappa=\lambda\pi$, we use this estimate in the following
\begin{eqnarray}
\nonumber
q_{II}[\Phi]&=&\Bigl{\|}{\partial \Phi \over \partial
x}\Bigr{\|}^2_{L^2(\Omega_{II})} + \Bigl{\|}{\partial \Phi \over \partial
y}\Bigr{\|}^2_{L^2(\Omega_{II})} - {\pi^2\over
4d^2}\,\|\Phi\|^2_{L^2(\Omega_{II})} \geq\\
\nonumber\\
\nonumber
&\geq& \Bigl{\|}{\partial \Phi \over \partial x}\Bigr{\|}^2_{L^2
(\Omega_{II})}\bigl (1-\kappa^2\bigr
)-\sqrt{2\delta}{\kappa\pi\over 2d}\,\bigl{\|}\,|\Phi_-|+|\Phi_+|\,\bigr
{\|}_{L^2(0,d)} \Bigl{\|}{\partial \Phi \over \partial
x}\Bigr{\|}_{L^2(\Omega_{II})}-\\
\nonumber\\
\label{qII1}
&-&{\kappa\pi\over
2d}\,\|\Phi_-\|_{L^2(0,d)}\,\|\Phi_+\|_{L^2(0,d)}.
\end{eqnarray}
Let us assume that there exists $\Phi \in \mathcal{Q}(\Omega)$ such
that $q[\Phi]<0$.
Taking into consideration (\ref{qI+qIII})
we conclude that $q_{II}[\Phi]<0$ for considered $\Phi\in \mathcal{Q}$.
Then also
$$
\Bigl{\|}{\partial \Phi \over \partial x}\Bigr{\|}^2_{L^2
(\Omega_{II})}\bigl (1-\kappa^2\bigr
)-\sqrt{2\delta}{\kappa\pi\over 2d}\,\bigl{\|}\,|\Phi_-|+|\Phi_+|\,\bigr
{\|}_{L^2(0,d)} \Bigl{\|}{\partial \Phi \over \partial
x}\Bigr{\|}_{L^2(\Omega_{II})}-{\kappa\pi\over 2d}\,\|\Phi_-\|_{L^2(0,d)}\,
\|\Phi_+\|_{L^2(0,d)}<0.
$$
This inequality holds obviously for $\kappa\geq 1$.
Let us solve this inequality for $\kappa<1$.
Then using the triangle inequality together with the inequality
\begin{equation}\label{pomoc1}
2\|\Phi_-\|_{L^2(0,d)}\|\Phi_+\|_{L^2(0,d)} \leq {1\over2}\,\bigl
(\|\Phi_-\|_{L^2(0,d)}+\|\Phi_+\|_{L^2(0,d)}\bigr )^2
\end{equation}
we get
\begin{equation}\label{Phi'}
\Bigl{\|}{\partial \Phi \over \partial x}\Bigr{\|}_{L^2
(\Omega_{II})}<{\kappa\over 2\sqrt{2\delta}(1-\kappa)}\,
\Bigl (\|\Phi_-\|_{L^2(0,d)}+\|\Phi_+\|_{L^2(0,d)}\Bigr
).
\end{equation}
We use this result in the following estimate
\begin{eqnarray}\nonumber
\|\Phi_+-\Phi_-\|_{L^2(0,d)}&=& \sqrt {\int_0^d\Bigl | \int_{-\delta}^\delta {\partial \Phi \over \partial
x}\D x \Bigr |^2 \D y} \leq \sqrt{2\delta}\Bigl{\|}{\partial \Phi \over \partial x}\Bigr{\|}_{L^2
(\Omega_{II})}<\\
\nonumber\\
\label{Phi+-Phi-}
&<&{\kappa\over 2(1-\kappa)}\,\Bigl (\|\Phi_-\|_{L^2(0,d)}+\|\Phi_+\|_{L^2(0,d)}\Bigr
).
\end{eqnarray}
Now we use estimates (\ref{qI+qIII}),(\ref{qII1}) together with
inequality (\ref{pomoc1}), the triangle inequality and
\begin{equation}\label{pomoc2}
\bigl (\|\Phi_-\|_{L^2(0,d)}+\|\Phi_+\|_{L^2(0,d)}\bigr )^2 \leq
2\bigl (\|\Phi_-\|^2_{L^2(0,d)}+\|\Phi_+\|^2_{L^2(0,d)}\bigr )
\end{equation}
to estimate whole functional $q[\Phi]$.
Denoting $A_0=a_0(-\delta)$, $B_0=b_0(\delta)$ we obtain
\begin{eqnarray}\nonumber
q[\Phi]&\geq&{\pi\over d}\,\sum_{k=1}^\infty\sqrt{k(k+1)} \bigl
(|A_k|^2+|B_k|^2\bigr )+\Bigl{\|}{\partial \Phi \over \partial x}\Bigr{\|}^2_{L^2
(\Omega_{II})}\bigl (1-\kappa^2\bigr
)-\\
\nonumber\\
\nonumber
&-&\sqrt{2\delta}{\kappa\pi\over 2d}\,\bigl{\|}\,|\Phi_-|+|\Phi_+|\,\bigr
{\|}_{L^2(0,d)} \Bigl{\|}{\partial \Phi \over \partial
x}\Bigr{\|}_{L^2(\Omega_{II})}-{\kappa\pi\over 2d}\,\|\Phi_-\|_{L^2(0,d)}\,
\|\Phi_+\|_{L^2(0,d)}\geq\\
\nonumber\\
\nonumber
&\geq&{\pi\sqrt2\over d}\,\sum_{k=1}^\infty \bigl (|A_k|^2+|B_k|^2\bigr
)+\\
\nonumber\\
\nonumber
&+&(1-\kappa^2)\biggl (\Bigl{\|}{\partial \Phi \over \partial x}\Bigr{\|}_{L^2
(\Omega_{II})}-\sqrt{2\delta}{\kappa\pi\over 4d}\,{\bigl{\|}\,|\Phi_-|+|\Phi_+|\,\bigr
{\|}_{L^2(0,d)}\over 1-\kappa^2}\biggr)^2-\\
\nonumber\\
\nonumber
&-&{\kappa^3\pi\over 8 d}\,{\bigl{\|}\,|\Phi_-|+|\Phi_+|\,\bigr
{\|}^2_{L^2(0,d)}\over (1-\kappa^2)}-{\kappa\pi\over 4 d}\, \bigl
(\|\Phi_-\|^2_{L^2(0,d)}+\|\Phi_+\|^2_{L^2(0,d)}\bigr)\geq\\
\nonumber\\
\label{correction}
&\geq&{\pi\sqrt2\over d}\,\sum_{k=1}^\infty \bigl (|A_k|^2+|B_k|^2\bigr
)-{\kappa\pi\over 4 d(1-\kappa^2)}\sum_{k=0}^\infty \bigl (|A_k|^2+|B_k|^2\bigr
),
\end{eqnarray}
where we used also (\ref{PhiI}) and (\ref{PhiIII}) together with (\ref{Ak}) and
(\ref{Bk}).
We assumed $q[\Phi]<0$, thus the same is true for the last
expression in (\ref{correction}) and so we obtain for $\kappa<{\sqrt{129}-1\over 8\sqrt2}\doteq
0.92$
\begin{equation}\label{A0^2+B0^2}
\sum_{k=1}^\infty \bigl (|A_k|^2+|B_k|^2\bigr )< {\kappa\over4\sqrt2(1-\kappa^2)-
\kappa}\bigl (|A_0|^2+|B_0|^2\bigr ).
\end{equation}
Now we need estimates (\ref{pomoc2}), (\ref{Phi+-Phi-}) and (\ref{A0^2+B0^2})
together with known inequalities for nonnegative numbers
\begin{eqnarray}\nonumber
2ab&\leq& a^2+b^2\\
\nonumber (a+b+c)^2&\leq& 3 (a^2+b^2+c^2)
\end{eqnarray}
and the triangle inequality to do the last step of the proof.
First notice, that
\begin{equation}\label{konec1}
{2\over d}\,\Bigl \|A_0\sin{\pi y\over 2 d}-B_0\cos{\pi y\over 2
d}\Bigr \|^2_{L^2(0,d)}=|A_0|^2+|B_0|^2-{4\over\pi}\,\Re(A_0 B_0)\geq
\Bigl (1-{2\over\pi}\Bigr)\bigl(|A_0|^2+|B_0|^2\bigr).
\end{equation}
On the other hand
\begin{eqnarray}\nonumber
&\,&{2\over d}\,\Bigl \|A_0\sin{\pi y\over 2 d}-B_0\cos{\pi y\over 2
d}\Bigr \|^2_{L^2(0,d)}=\\
\nonumber\\
\nonumber
&=&\Bigl \|\Phi_--\Phi_+-\sqrt{2\over
d}\,\sum_{k=1}^\infty\Bigl (A_k\sin{(2k+1)\pi y\over 2 d}-B_k\cos{(2k+1)\pi y\over 2
d}\Bigr ) \Bigr\|^2_{L^2(0,d)}\leq\\
\nonumber\\
\nonumber
&\leq&\Biggl
(\bigl\|\Phi_--\Phi_+\bigr\|_{L^2(0,d)}+\sqrt{\sum_{k=1}^\infty
|A_k|^2} + \sqrt{\sum_{k=1}^\infty|B_k|^2}\, \Biggr )^2\leq\\
\nonumber\\
\nonumber
&\leq& \Biggl ({\kappa\over2(1-\kappa)}\Bigl (
\|\Phi_-\|_{L^2(0,d)}+\|\Phi_+\|_{L^2(0,d)}\Bigr) + \sqrt{\sum_{k=1}^\infty
|A_k|^2} + \sqrt{\sum_{k=1}^\infty|B_k|^2}\, \Biggr )^2<\\
\nonumber\\
\nonumber
&<& 3 \Biggl ({\kappa^2\over2(1-\kappa)^2}\, \sum_{k=0}^\infty
\Bigl (|A_k|^2+|B_k|^2\Bigr )+ \sum_{k=1}^\infty
\Bigl (|A_k|^2+|B_k|^2\Bigr ) \Biggr )<\\
\nonumber\\
\nonumber
&<&3{2\sqrt2(1+\kappa)\kappa^2+\kappa(1-\kappa)\over
(1-\kappa)\bigl(4\sqrt2( 1-\kappa^2) -\kappa \bigr )}\, \bigl
(|A_0|^2+|B_0|^2\bigr).
\end{eqnarray}
We compare this inequality with (\ref{konec1}) and we have
\begin{equation}\label{konec2}
1-{2\over \pi}\,<3\kappa{2\sqrt2(1+\kappa)\kappa+1-\kappa\over
(1-\kappa)\bigl(4\sqrt2( 1-\kappa^2) -\kappa \bigr )}.
\end{equation}
We recall, that we investigate this inequality for $0<\kappa<{\sqrt{129}-1\over
8\sqrt2}$.
The right hand side of (\ref{konec2}) is a positive, continuous function of
$\kappa$ on this interval and it goes to zero as $\kappa$ does.
Hence there must exist a number $\kappa_0$ (and so $\Lambda_1$),
such that this inequality does no hold for
$\kappa\in(0,\kappa_0)$, resp. $\lambda \in (0,\Lambda_1)$.
Thus $q[\Phi]\geq 0$ for all $\Phi \in \mathcal{Q}(\Omega)$ for these
values of the parameter $\lambda$, which has been to prove.
\qed
\end{PF}
\begin{rem}
We also know that the discrete eigenvalues emerge at the essential spectrum
treshold and they are continuous functions of $\lambda$ in both
our specific cases.
The rigorous formulation and the proof of this statement are given
in Appendix A.
\end{rem}

\section{Numerical results}

We have solved the Schr\"odinger equation corresponding to our
systems numerically.
Since $\Omega$ consists of three rectangular regions, the easiest
way to do that is the mode-matching method (see \eg\ \cite{ESTV}).
The results are shown in Figures~2--5.
On the Figures~4 and 5 we can notice that the first eigenvalue in
the model A appears for $\Lambda_0 >0$.
Our numerical results indicate that $\Lambda_0\doteq0.26$ in agreement with our analytical proofs.
Solving numerically the equation $q_{II}[\Phi_\sigma]=0$ with right
hand side given by (\ref{value qII}) we get the result
$\Lambda_2\doteq0.34$.
In the same way from (\ref{konec2}) we get after numerical computation
$\Lambda_1\doteq0.08$.
Hence $\Lambda_1<\Lambda_0<\Lambda_2$ which we have expected.
\begin{figure}
\begin{center}
\includegraphics[height=10cm, width=15cm]{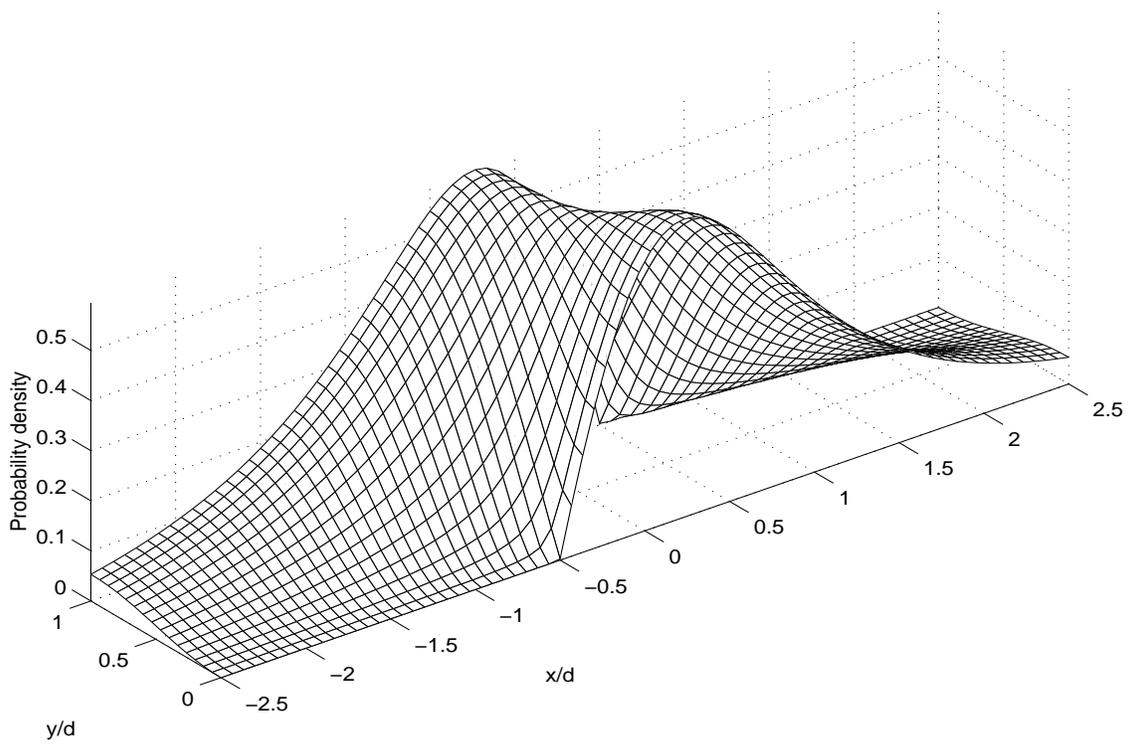}
\end{center}
\caption{Probability density $|\psi|^2$ (in units of $d^{-2}$) for
the bound state in model A, $\lambda={1\over2}$, coordinates
$x,\, y$ in the units of $d$. Only one discrete bound state exists
here.}
\end{figure}
\begin{figure}
\begin{center}
\includegraphics[height=9cm, width=15cm]{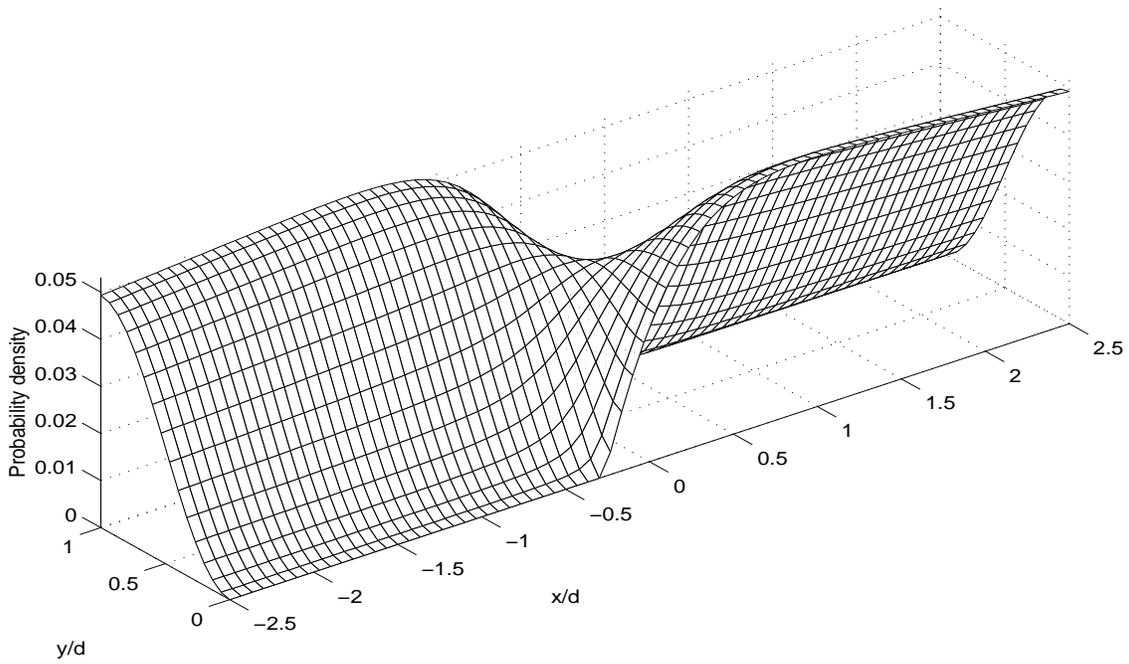}
\end{center}
\caption{The same as in Fig.2 but $\lambda=0.27$. It is close to
the treshold for bound state appearance.}
\end{figure}
\begin{figure}
\begin{center}
\includegraphics[height=9cm, width=15cm]{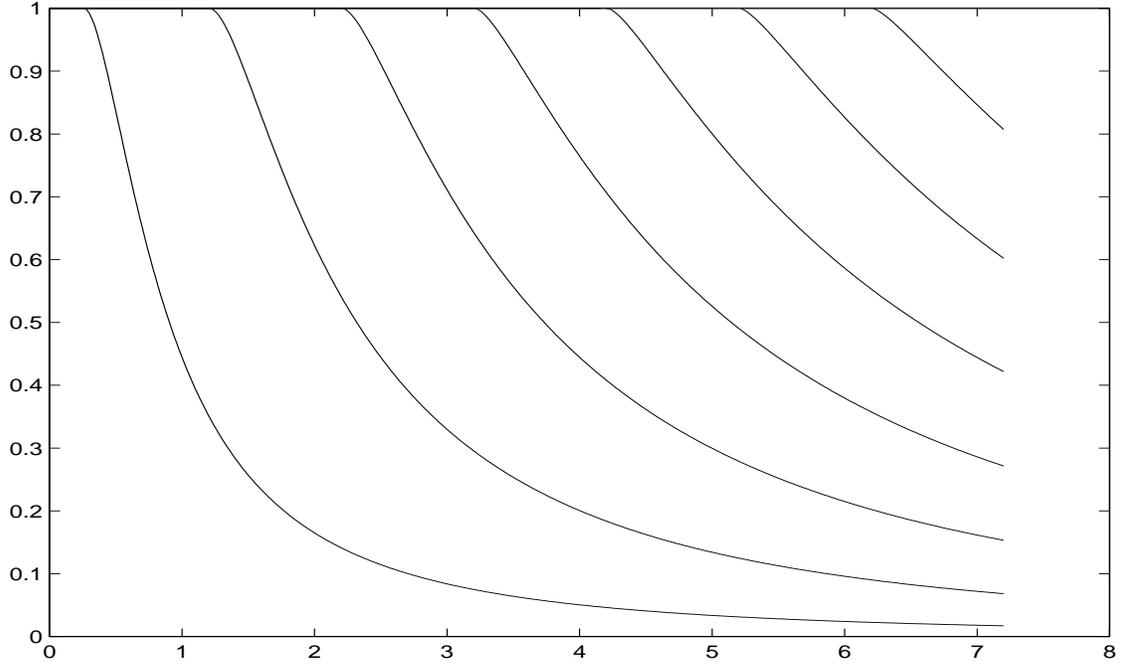}
\end{center}
\caption{Eigenvalues (in the units of $\mu={\pi^2\over4 d^2}$) for
the model A in dependence on $\lambda$.}
\end{figure}
\begin{figure}
\begin{center}
\includegraphics[height=9cm, width=15cm]{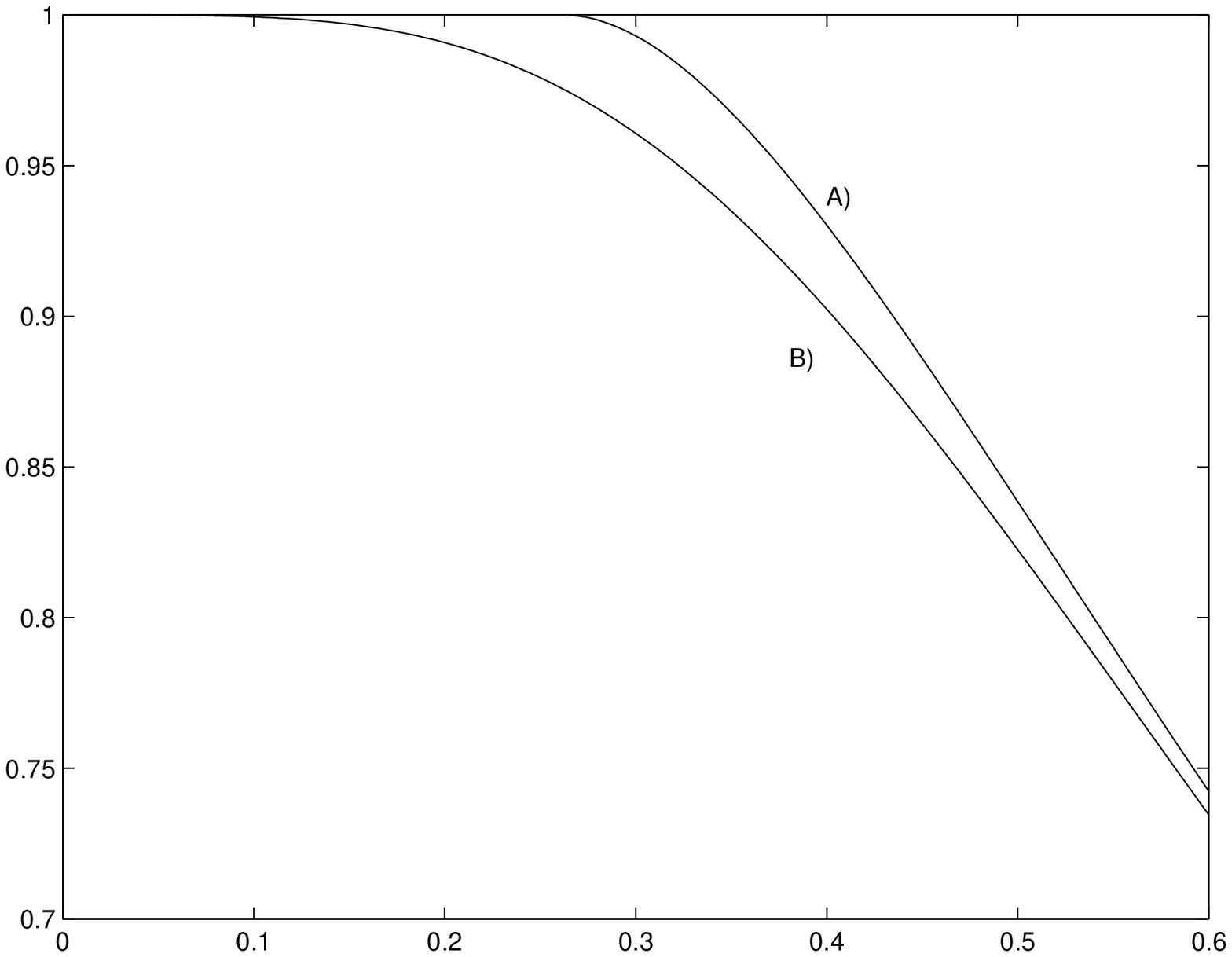}
\end{center}
\caption{First eigenvalues for the model A and model B in
dependence of $\lambda$. While in model B eigenvalue exists for
any $\lambda>0$ in model A appears at $\Lambda>\Lambda_0$ (numerical
calculations indicate $\Lambda_0\doteq0.26$).}
\end{figure}
\section*{Acknowledgements}
The authors thank Professor Pavel Exner for discussions on the
problem formulation and expected results.
The work is supported by GA ASCR grant IAA 1048101.

\appendix
\renewcommand{\theequation}{\Alph{section}\arabic{equation}}
\renewcommand{\thesection}{Appendix \Alph{section}:}
\section{The continuity of the eigenvalues}
\setcounter{equation}{0}

First we define some notations for purposes of this appendix.
Let $\lambda>0$ then
$$
Q(\lambda)=\bigl\{ f\in H^1(\Omega)\,|\, f\upharpoonright
\mathcal{D}(\lambda)=0 \bigr \},
$$
where
\begin{eqnarray}
\nonumber
\mathcal{D}(\lambda)&=&\Bigl{\{}\langle x,0\rangle \,|\, x<-\lambda d
\Bigr{\}} \cup \Bigl{\{}\langle x,d\rangle | x>\lambda d
\Bigr{\}}\\
\nonumber \textrm{or}\\
\nonumber
\mathcal{D}(\lambda)&=& \Bigl{\{}\langle x,d\rangle \,|\, (x<-\lambda d) \vee
 (x>\lambda d)\Bigr{\}}.
\end{eqnarray}
So we have stressed only the dependence of the form domain on the
parameter $\lambda$.
We define
\begin{eqnarray}\nonumber
U(\varphi_1,\ldots,\varphi_m;\lambda)&=&\inf_{\begin{array}{c}
\psi \in Q(\lambda),\psi \neq 0\\
\psi\bot\varphi_1,\ldots,\varphi_m
\end{array}}
{\|\nabla \psi \|^2_{L^2(\Omega)} \over \|\psi
\|^2_{L^2(\Omega)}}\\
\nonumber\\
\nonumber \mu_n(\lambda)&=&\sup_{\varphi_1,\ldots,\varphi_{n-1} \in
L^2(\Omega)} U(\varphi_1,\ldots,\varphi_{n-1};\lambda).
\end{eqnarray}
From minimax principle \cite{RS}, we know that for every $n=1,2,\ldots$,
$\mu_n(\lambda)$ is either the $n-$th eigenvalue of the operator
(counting multiplicity) or the bottom of its essential spectrum.
The aim of this appendix is to show that $\mu_n(\lambda)$ are
continuous functions in $(0,\infty)$.
\begin{lemma}
Functions $\mu_n\,:\,\lambda \mapsto \mu_n(\lambda)$ are
nonincreasing, finite and continuous in $(0,\infty)$ for every
$n=1,2,\ldots$.
\end{lemma}
\begin{PF}
We know from the minimax principle and the Dirichlet-Neumann bracketing
\cite{RS} that $0\leq \mu_n(\lambda) \leq {\pi^2\over 4 d^2}$ for all
$\lambda>0$ and $n=1,2,\ldots$.
Let $\lambda_1>\lambda_2$.
Then $Q(\lambda_2) \subset Q(\lambda_1)$ and so
for any $m=0,1,\ldots$ and any $m$-tuple $\varphi_1,\ldots,\varphi_m$
$$
U(\varphi_1,\ldots,\varphi_m;\lambda_1)\leq
U(\varphi_1,\ldots,\varphi_m;\lambda_2)\leq \mu_{m+1}(\lambda_2).
$$
Because this inequalities hold for every $m$-tuple
$\varphi_1,\ldots,\varphi_m$ from our Hilbert space they must be
fulfilled even for supremum over these $m$-tuples.
Hence
\begin{equation}\label{monotonnost mu}
\mu_{m+1}(\lambda_1)\leq \mu_{m+1}(\lambda_2)
\end{equation}
for every $m=0,1,\ldots$ and $\mu_{m+1}$ are thus nonincreasing.

For any $\varrho>1$ we define
$\varphi^{(\varrho)}(x,y)=\varphi(\varrho x,y)$.
The equivalences $\varphi\in Q(\lambda) \Leftrightarrow \varphi^{(\varrho)}\in
Q({\lambda\over \varrho})$ and $\varphi \bot \psi \Leftrightarrow \varphi^{(\varrho)}
\bot \psi^{(\varrho)}$ are obvious.
Moreover
\begin{equation}\label{vypocet q}
{\|\nabla \varphi^{(\varrho)} \|^2_{L^2(\Omega)} \over
\|\varphi^{(\varrho)}
\|^2_{L^2(\Omega)}}= {\varrho^2\Bigr\|{\partial \varphi \over\partial
x}\Bigr\|^2_{L^2(\Omega)}+\Bigr\|{\partial \varphi \over\partial
y}\Bigr\|^2_{L^2(\Omega)} \over \|\varphi
\|^2_{L^2(\Omega)}} \leq
\varrho^2 {\|\nabla \varphi \|^2_{L^2(\Omega)} \over \|\varphi
\|^2_{L^2(\Omega)}}.
\end{equation}
Let integer $m\geq 0$, $\lambda>0$ and
$\varphi_1,\ldots,\varphi_m\in L^2(\Omega)$ are chosen arbitrarily.
Then for any $\varphi\in Q(\lambda)$ such that $\varphi \bot
\varphi_1,\ldots,\varphi_m$ we know $\varphi^{(\varrho)}_1,\ldots,\varphi^{(\varrho)}_m\in
L^2(\Omega)$, $\varphi^{(\varrho)}\in Q({\lambda\over \varrho})$ and  $\varphi^{(\varrho)} \bot
\varphi^{(\varrho)}_1,\ldots,\varphi^{(\varrho)}_m$.
Then using (\ref{vypocet q}) we get
$$
U\Bigl(\varphi^{(\varrho)}_1,\ldots,\varphi^{(\varrho)}_m;{\lambda\over\varrho}\Bigr)
\leq {\|\nabla \varphi^{(\varrho)} \|^2_{L^2(\Omega)} \over
\|\varphi^{(\varrho)} \|^2_{L^2(\Omega)}} \leq\varrho^2 {\|\nabla \varphi \|^2_{L^2(\Omega)} \over \|\varphi
\|^2_{L^2(\Omega)}}.
$$
Because these inequalities hold for arbitrary $m$-tuple
$\varphi_1,\ldots\varphi_m$ it must hold even for the infimum of
${\|\nabla \varphi \|^2_{L^2(\Omega)} \over \|\varphi
\|^2_{L^2(\Omega)}}$ and
hence
$$
U\Bigl(\varphi^{(\varrho)}_1,\ldots,\varphi^{(\varrho)}_m;{\lambda\over\varrho}\Bigr)
\leq \varrho^2 U(\varphi_1,\ldots,\varphi_m;\lambda)
\leq\varrho^2 \mu_{m+1}(\lambda).
$$
Passing again to the supremum over all
$m$-tuples $\varphi^{(\varrho)}_1,\ldots,\varphi^{(\varrho)}_m$ and
taking into account (\ref{monotonnost mu}) we obtain
\begin{equation}\label{prvni odhad mu}
\mu_{m+1}(\lambda)\leq
\mu_{m+1}\Bigl({\lambda\over\varrho}\Bigr)\leq
\varrho^2\mu_{m+1}(\lambda),
\end{equation}
because $\lambda>{\lambda\over\varrho}$.
From the second inequality of (\ref{prvni odhad mu}) and from
(\ref{monotonnost mu}) we get
\begin{equation}\label{druhy odhad mu}
{1\over\varrho^2}\mu_{m+1}(\lambda)\leq
\mu_{m+1}(\lambda\varrho)\leq
\mu_{m+1}(\lambda)
\end{equation}
using $\lambda\varrho>\lambda$.
We recall that these inequalities hold for any
$m\in\{0,1,\ldots\}$.

For any $0<\lambda'<\lambda$ we set
$\varrho={\lambda\over\lambda'}$ and we put it to (\ref{prvni
odhad mu})
$$
\mu_{m+1}(\lambda)\leq\mu_{m+1}(\lambda')\leq \Bigl
({\lambda\over\lambda'}\Bigr)^2\mu_{m+1}(\lambda).
$$
We see that
$\lim_{\lambda'\rightarrow\lambda^-}\mu_{m+1}(\lambda')=\mu_{m+1}(\lambda)$.
We repeat the same procedure for $0<\lambda<\lambda'$ using
the inequalities (\ref{druhy odhad mu}) instead of (\ref{prvni
odhad mu}) and we get
$\lim_{\lambda'\rightarrow\lambda^+}\mu_{m+1}(\lambda')=\mu_{m+1}(\lambda)$,
which finishes the proof.
\qed
\end{PF}

\section{The dense set in $Q^A(q_0)$}
\setcounter{equation}{0}

The goal of the appendix is to prove the following lemma.
\begin{lemma}
The set $\tilde Q(\Omega)=\bigl \{\psi \in
C^\infty(\overline\Omega)\,|\,\psi=\tilde\psi\upharpoonright\Omega,\,
\tilde\psi \in C_0^\infty(\mathbb{R}^2),\psi\upharpoonright
\mathcal{D}=0 \bigr \} $ is dense in the set $Q^A(q_0)$ with
respect to the $H^1(\Omega)$ norm.
\end{lemma}
\begin{PF}
Let $\varphi\in Q^A(q_0)$.
Then for every $\eps>0$ we can find the function $\varphi_1^{(\eps)}$
from the set $\bigl \{\psi \in
C^\infty(\overline\Omega)\,|\,\psi=\tilde\psi\upharpoonright\Omega,\,
\tilde\psi \in C_0^\infty(\mathbb{R}^2) \bigr \}$ such that
$\|\varphi-\varphi_1^{(\eps)}\|_{H^1(\Omega)}<\eps$ (see \eg\ \cite{Adams},
Theorem 3.18).
Due to the Theorem 5.22 in \cite{Adams} we can also write
$\|(\varphi-\varphi_1^{(\eps)})\upharpoonright \mathcal{D}\|_{L^2(
\mathcal{D})}<\sqrt{C}\eps$, where $C$ is a constant.
Let $\gamma$ be the function from the class $C^\infty(\mathbb{R})$
with the following properties: $0\leq\gamma(t)\leq1$ for $t\in
\mathbb{R}$, $\gamma(t)=0$ for $t\leq {1\over2}$, $\gamma(t)=1$ for
$t\geq 1$ and $\bigl |{d\gamma(t)\over d t}\bigr |\leq {\Gamma\over 5}$
for some constant $\Gamma$ and every $t\in\mathbb{R}$.
For $\langle x,y\rangle \in \Omega$ and $\eps<d/2$ we define
\begin{eqnarray}\nonumber
\omega_\eps^{(1)}(x,y)&=&\left \{ \begin{array}{lcl}
\gamma\Bigl({4(d-y)\over \eps}\Bigr)\gamma\Bigl({\sqrt{(x-\delta)^2+(y-d)^2}\over \eps}\Bigr)
& \textrm{for} & x\geq\delta\\
\\
\gamma\Bigl({\sqrt{(x-\delta)^2+(y-d)^2}\over \eps}\Bigr) &
\textrm{for} &x\leq \delta
\end{array}\right . ,\\
\nonumber\\
\nonumber
\omega_\eps^{(2)}(x,y)&=&\left \{ \begin{array}{lcl}
\gamma\Bigl({4y\over \eps}\Bigr)\gamma\Bigl({\sqrt{(x+\delta)^2+y^2}\over \eps}\Bigr)
& \textrm{for} & x\leq -\delta\\
\\
\gamma\Bigl({\sqrt{(x+\delta)^2+y^2}\over \eps}\Bigr) &
\textrm{for} &x\geq -\delta
\end{array}\right .
\end{eqnarray}
and $\omega_\eps=\omega_\eps^{(1)}\omega_\eps^{(2)}$.
Obviously $\varphi_2^{(\eps)}=\varphi_1^{(\eps)}\omega_\eps \in
\tilde Q(\Omega)$.
Let we denote
\begin{eqnarray}\nonumber
\Omega_\eps&=&\Bigl \{\langle x,y\rangle \in
\Omega\,|\,\textrm{dist}\bigl(\langle x,y\rangle, \mathcal{D}
\bigr)<\eps\Bigr \}\\\nonumber
\Omega_\eps^1&=&\Bigl((\delta,\infty)\times (d-\eps,d)\Bigr)\cup
\Bigl((-\infty,-\delta)\times (0,\eps)\Bigr)\\\nonumber
\Omega_\eps^2&=&\Bigl((\delta-\eps,\delta)\times (d-\eps,d)\Bigr)\cup
\Bigl((-\delta,-\delta+\eps)\times (0,\eps)\Bigr)\\\nonumber
\Omega_\eps^3&=&\Bigl((\delta-\eps,\delta+\eps)\times (d-\eps,d)\Bigr)\cup
\Bigl((-\delta-\eps,-\delta+\eps)\times (0,\eps)\Bigr).
\end{eqnarray}
Now we are going to estimate
$\|\varphi_2^{(\eps)}-\varphi\|^2_{H^1(\Omega)}$.
\begin{eqnarray}\nonumber
\|\varphi_2^{(\eps)}-\varphi\|^2_{H^1(\Omega)}&=&
\|(\varphi_1^{(\eps)}-\varphi)\omega_\eps+\varphi(\omega_\eps-1)
\|^2_{L^2(\Omega)}+\\
\nonumber\\
\nonumber
&+&\|\nabla(\varphi_1^{(\eps)}-\varphi)\omega_\eps+\nabla \varphi(\omega_\eps-1)
+\varphi_1^{(\eps)}\nabla\omega_\eps\|^2_{L^2(\Omega)}\leq\\
\nonumber\\
\nonumber
&\leq&3\biggl(
\|\varphi_1^{(\eps)}-\varphi\|^2_{H^1(\Omega)}+\|\varphi\|^2_{H^1(\Omega_\eps)}
+ {\Gamma^2 \over \eps^2}
\|\varphi_1^{(\eps)}\|^2_{L^2(\Omega_\eps)}\biggr)
\end{eqnarray}
Now we continue by estimating of
$\|\varphi_1^{(\eps)}\|^2_{L^2(\Omega_\eps)}$.
\begin{eqnarray}\nonumber
\|\varphi_1^{(\eps)}\|^2_{L^2(\Omega_\eps)}&\leq&\|\varphi_1^{(\eps)}\|^2_{L^2(\Omega^1_\eps)}
+ \|\varphi_1^{(\eps)}\|^2_{L^2(\Omega^2_\eps)}\\
\nonumber\\\nonumber
\|\varphi_1^{(\eps)}\|^2_{L^2(\Omega^1_\eps)}&=& \int_\delta^\infty
\int_{d-\eps}^d |\varphi_1^{(\eps)}(x,y)|^2\D y\D x
+\int^{-\delta}_{-\infty} \int_0^\eps |\varphi_1^{(\eps)}(x,y)|^2\D y\D x
\end{eqnarray}
Because the estimates of the second term in the previous
expression will be analogous to ones of the first term we will not
write the second integral, but only ``second term" instead of it.
\begin{eqnarray}\nonumber
\|\varphi_1^{(\eps)}\|^2_{L^2(\Omega^1_\eps)}&=& \int_\delta^\infty
\int_{d-\eps}^d \bigl |\varphi_1^{(\eps)}(x,d)-\int_y^d{\partial
\varphi_1^{(\eps)}\over \partial t}(x,t)\D t \bigr |^2\D y\D x +
\textrm{second term}\leq\\
\nonumber\\\nonumber
&\leq&2\int_\delta^\infty \int_{d-\eps}^d \biggl ( \bigl
|\varphi_1^{(\eps)}(x,d)\bigr |^2 + \eps \int_{d-\eps}^d \Bigl |{\partial
\varphi_1^{(\eps)}\over \partial t}(x,t)\Bigr |^2\D t\biggr )\D
y\D x + \textrm{second term}=\\
\nonumber\\\nonumber
&=&2\biggl(\eps\|\varphi_1^{(\eps)}\upharpoonright
\mathcal{D}\|^2_{L^2( \mathcal{D})} + \eps^2 \Bigl \|{\partial
\varphi_1^{(\eps)}\over \partial
y}\Bigr\|^2_{L^2(\Omega_\eps^1)}\biggr)\leq 2\eps^2\bigl( C\eps
+\|\varphi_1^{(\eps)}\|^2_{H^1(\Omega_\eps^1)}\bigr)\leq\\
\nonumber\\\nonumber
&\leq&2\eps^2\Bigl( C\eps + 2\bigl(\|\varphi_1^{(\eps)}-\varphi
\|^2_{H^1(\Omega)}+\|\varphi\|^2_{H^1(\Omega_\eps^1)}
\bigr)\Bigr)\leq 2\eps^2\Bigl( C\eps + 2\eps^2+2
\|\varphi\|^2_{H^1(\Omega_\eps^1)}\Bigr)
\end{eqnarray}
For any $z\in (\delta,\delta+\eps)$ we calculate
\begin{eqnarray}\nonumber
\|\varphi_1^{(\eps)}\|^2_{L^2(\Omega^2_\eps)}&=&
\int_{\delta-\eps}^\delta
\int_{d-\eps}^d \bigl |\int_x^z{\partial
\varphi_1^{(\eps)}\over \partial t}(t,y)\D t+ \int_y^d{\partial
\varphi_1^{(\eps)}\over \partial s}(z,s)\D s -\varphi_1^{(\eps)}(z,d)
\bigr |^2\D y\D x +\textrm{s.t.}\leq\\
\nonumber\\\nonumber
&\leq&3\int_{\delta-\eps}^\delta\int_{d-\eps}^d \biggl( 2\eps
\int_{\delta-\eps}^{\delta+\eps}\Bigl |{\partial
\varphi_1^{(\eps)}\over \partial t}(t,y)\Bigr |^2 \D t+\eps
\int_{d-\eps}^d \Bigl |{\partial
\varphi_1^{(\eps)}\over \partial s}(z,s)\Bigr |^2 \D s +\\
\nonumber\\\nonumber
&+&\bigl |\varphi_1^{(\eps)}(z,d)\bigr |^2\biggr )\D y\D x +
\textrm{second term}.
\end{eqnarray}
When we integrate the last inequality over $z$ from $\delta$ to
$\delta+\eps$ we obtain
$$
\eps\|\varphi_1^{(\eps)}\|^2_{L^2(\Omega^2_\eps)}\leq 3 \biggl (
2\eps^3\Bigl\|{\partial \varphi_1^{(\eps)}\over \partial x}\Bigr
\|^2_{L^2(\Omega_\eps^3)} +\eps^3\Bigl\|{\partial \varphi_1^{(\eps)}\over \partial
y}\Bigr \|^2_{L^2(\Omega_\eps^3)}+\eps^2 \bigl\|\varphi_1^{(\eps)}
\upharpoonright \mathcal{D}\bigr\|^2_{L^2(\mathcal{D})}\biggr)
$$
and so
$$
\|\varphi_1^{(\eps)}\|^2_{L^2(\Omega^2_\eps)}\leq 6\eps^2
\|\varphi_1^{(\eps)}\|^2_{H^1(\Omega^3_\eps)}+3C\eps^3 \leq
12\eps^2\Bigl (\eps^2+\|\varphi\|^2_{H^1(\Omega^3_\eps)}\Bigr
)+3C\eps^3.
$$
We give both these estimates together and we get
$$
\|\varphi_1^{(\eps)}\|^2_{L^2(\Omega_\eps)}\leq 5 C\eps^3 +16
\eps^4+16\eps^2
\|\varphi\|^2_{H^1(\Omega^1_\eps\cup\Omega^2_\eps)}.
$$
Hence
$$
\|\varphi_2^{(\eps)}-\varphi\|^2_{H^1(\Omega)}\leq 3 \biggl (
(1+16\Gamma^2)(\eps^2+\|\varphi\|^2_{H^1(\Omega^1_\eps\cup
\Omega^2_\eps)}) + 5C\Gamma^2\eps \biggr).
$$
Here $\lim_{\eps\rightarrow0^+}\|\varphi\|^2_{H^1(\Omega_\eps^1
\cup \Omega_\eps^2)}=0$, \eg\ by the dominated convergence.
So $\|\varphi_2^{(\eps)}-\varphi\|^2_{H^1(\Omega)}$ goes to zero
as $\eps$ does which has been to show.
\qed
\end{PF}
\begin{rem}
Our set $\tilde Q$ is obviously a subset of the set
$\mathcal{Q}$ defined in the proof of the Theorem 3.
Hence $\mathcal{Q}$ is a dense set in $Q^A(q_0)$ as well.
\end{rem}

\end{document}